\documentclass[times, twocolumn]{aastex62}
\usepackage{graphicx}
\usepackage{amssymb}
\usepackage{mathrsfs} 
\usepackage{natbib}
\usepackage{amsmath}
\usepackage[english]{babel}
\usepackage{url}
\usepackage{hyperref}
\usepackage{xcolor}
\usepackage{xspace}
\usepackage{ulem}

\newcommand{\haffet}{{\tt HAFFET}\xspace}        
\newcommand{\Ni}{$\rm ^{56}Ni$}

\newcommand{\Co}{$\rm ^{56}Co$}
\newcommand{\Fe}{$\rm ^{56}Fe$}
\newcommand{\Mni}{$M_{\rm Ni}$}
\newcommand{\HeI}{${\rm He~I}~\lambda5876$~}
\newcommand{\OI}{${\rm O~I}~\lambda7772$~}
\newcommand{\Ek}{$E_{\rm Kin}$}
\newcommand{\Mej}{$M_{\rm Ej}$}

\newcommand{\Msun}{{\ensuremath{\mathrm{M}_{\odot}}}}

\newcommand{\ang}{\,\AA~}

\newcommand{\github}[1]{\href{https://github.com/#1}{\includegraphics[width=10pt]{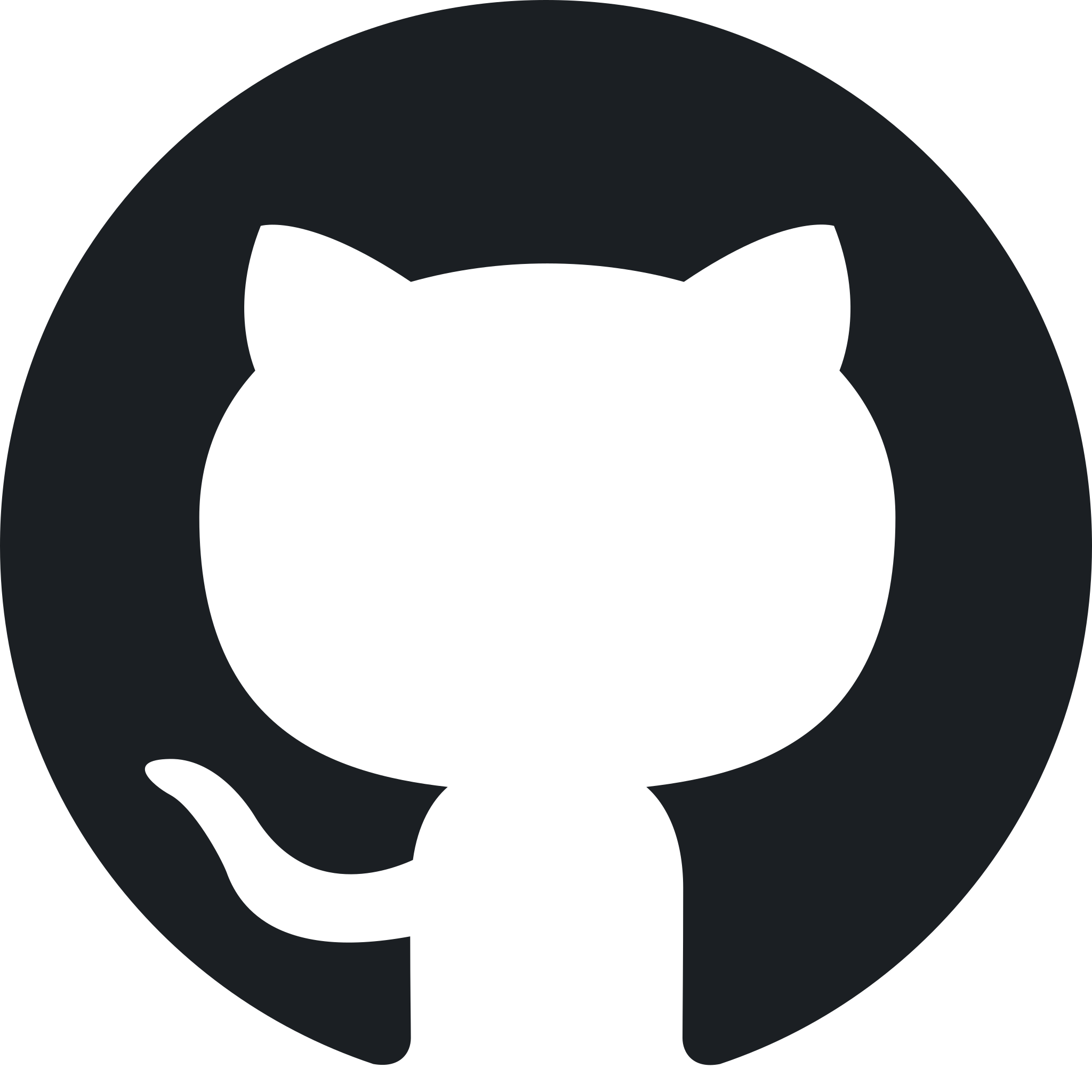}}}

\shorttitle{HAFFET}
\shortauthors{Yang et al.}

\begin{document}

\title{HAFFET: Hybrid Analytic Flux FittEr for Transients}

\author[0000-0002-2898-6532]{Sheng Yang}
\email{sheng.yang@hnas.ac.cn}
\affiliation{Department of Astronomy, The Oskar Klein Center, Stockholm University, AlbaNova, 10691 Stockholm, Sweden}
\affiliation{Henan Academy of Sciences, Zhengzhou 450046, Henan, China}
\author[0000-0003-1546-6615]{Jesper Sollerman}\affiliation{Department of Astronomy, The Oskar Klein Center, Stockholm University, AlbaNova, 10691 Stockholm, Sweden}

\begin{abstract}
The progenitors for many types of supernovae (SNe) are still unknown, and an approach to diagnose their physical origins is to investigate the light curve brightness and shape of a large set of SNe. However, it is often difficult to compare and contrast the existing sample studies 
due to differences in their approaches and assumptions, for example in how to eliminate host galaxy extinction, and this might lead to systematic errors when comparing the results.
We therefore introduce the Hybrid Analytic Flux FittEr for Transients (\haffet), a Python-based software package that can be applied to download photometric and spectroscopic data for transients from open online sources, derive bolometric light curves, and fit them to semi-analytical models for estimation of their physical parameters. In a companion study, we have investigated a large collection of SNe Ib and Ic observed with the Zwicky Transient Facility (ZTF) with \haffet, and here we detail the  methodology and the software package to encourage more users.
As large-scale surveys such as ZTF and LSST continue to discover increasing numbers of transients, tools such as \haffet will be critical for enabling rapid comparison of models against data in statistically consistent, comparable and reproducable ways. Additionally, \haffet is created with a Graphical User Interface mode, which we hope will boost the efficiency and make the usage much easier.
\github{saberyoung/HAFFET}
\end{abstract}


\section{Introduction}
\label{sec:intro}

Transients are celestial objects whose brightness varies on timescales from seconds to years, for example due to explosions of their progenitors. These objects help astronomers to investigate physics under extreme conditions, and there are currently a large number of surveys being carried out to search for extra-galactic transients by regularly monitoring the night sky. 
Detected objects include classical transients, such as supernovae (SNe) that are luminous optical transients from either a thermonuclear explosion of a white dwarf, or the core-collapse (CC) of a massive star, as well as new and exotic transients. e.g. the kilonova AT2017gfo \citep{GW170817MMA,sss17a,dlt17ck,vista,master,decam,lcogtkn, grawita, Smartt2017}. To correctly classify them, spectroscopy is required, however spectroscopic follow-up resources are scarce, and only a fraction of the currently discovered SNe can be classified in this way \citep{Kulkarni2018}.

Investigating the multi-band light curves (LCs) of the transients can also provide information about the mechanism that drives the explosions, such as shock cooling, circumstellar matter (CSM) interaction, radioactive decay or a central engine such as a magnetar. The information is encoded in the LCs in different aspects, e.g. their timescales, shapes and absolute luminosities, both for individual objects and for curated samples of specific types.
Much progress has been made by linking multi-wavelength light curves to models, both numerical hydrodynamical models and simpler semi-analytical estimates. For instance, for both SNe Ia and stripped envelope CC SNe, whose emission during the photospheric phase is normally powered by the radioactive decay of synthesized \Ni, \citet{arnett1982} made a semi-analytic model that can reconstruct the bolometric LCs for physical parameters of the explosion and the progenitors, such as the amount of ejected nickel mass, the ejecta mass and the kinetic energy. Therefore, exploring a large sample of SNe provides the opportunity to unveil their underlying physical origins.
Such analysis has been performed in a number of studies, e.g. \cite{barbarino, taddia2018csp, prentice2019, lyman16}, but the different studies often use different approaches and codes which makes the direct comparisons less straightforward. 
A generic code-package to handle light curve fitting for transients of different types, from different surveys, observed with varying cadences, is therefore useful to provide comparable results.
For this purpose, we present \haffet, a data-driven model fitter for transients. 
Often some spectroscopic information is also available for the SNe and needed to perform parts of the analysis, and \haffet also includes procedures for example for measuring ejecta velocities from spectra.

This paper aims to serve as a description of \haffet and its capabilities at the time of its first release.
This is complemented with a web-based
users guide, which the reader should always consult for more up-to-date information\footnote{\href{https://haffet.readthedocs.io/en/latest/}{https://haffet.readthedocs.io/en/latest/}}. This online documentation also contains more technical information and the code itself, whereas this paper aims to describe some of the possibilities in the code-package.
Furthermore, as a scientific proof-of-concept of \haffet, we are in a companion paper exploring a large sample of Type Ib and Ic SNe, observed by the Zwicky Transient Facility \citep[ZTF;][]{Bellm2019, Graham2019}, and spectroscopically classified as part of the Bright Transient Survey (BTS) \citep{fremling2020, perley2020}. That paper (Yang et al. in preparation) contains the analysis of the objects that are used here to demonstrate the capabilities of \haffet, and is 
hereafter referred to as Paper I. 
We have also estimated the luminosities for ZTF SNe Ic-BL using HAFFET in \cite{Corsi2022}, \cite{Gokul2023} and in \cite{Anand2023}.

This paper is organized as follows. 
In Section~\ref{sec:haffet} we present the code design of \haffet. In Section~\ref{sec:data}, we describe how \haffet handle meta and observational data, and in Section~\ref{sec:fitting} we discuss in detail the different available models designed for different types of observational data. 
The fitting process and outputs by which users can share their results are outlined in Section~\ref{sec:outputs}.
In Section~\ref{sec:application} we further explore the sample of ZTF SNe Ib and Ic with \haffet, complementary to Paper I. Finally, Section~\ref{sec:conclusion} presents our conclusions and provides a short discussion where we put our results in context.

\section{HAFFET}
\label{sec:haffet}

In this section, we present the code's design, workflow, and diverse operating modes within the framework of \haffet.

To begin, the code design of \haffet is meticulously crafted to offer flexibility and scalability, catering to various types of transient data and analytical necessities. It employs a modular structure, empowering users to selectively engage desired functionalities based on real-world scenarios.

\haffet offers three distinct methods of invocation: it can be employed as a Python package, executed as a standalone file, or accessed through a user-friendly Graphical User Interface (GUI). These approaches are functionally equivalent, all triggering the same core processes, and the outcomes they generate are also mutually interchangeable. The assortment of distinct operating modes is tailored to accommodate diverse analytical objectives. For instance, one mode may be tailored for light curve fitting and parameter estimation, while another could specialize in spectrum data processing and analysis. Users are empowered to select the most appropriate operating mode aligned with their specific analytical needs.

Moving on to the workflow, \haffet encompasses several key stages. Users initiate the process by providing observational data of the transient, encompassing light curves, spectra, and other pertinent information. Subsequently, users have the option to choose from a range of suitable operating modes for data analysis. The chosen operating mode guides \haffet to execute relevant data processing and fitting algorithms. Finally, users are able to access comprehensive analytical outcomes, including derived physical parameters and fitted light curve results.

\begin{figure*}
\centering
    \includegraphics[width=0.6\textwidth]{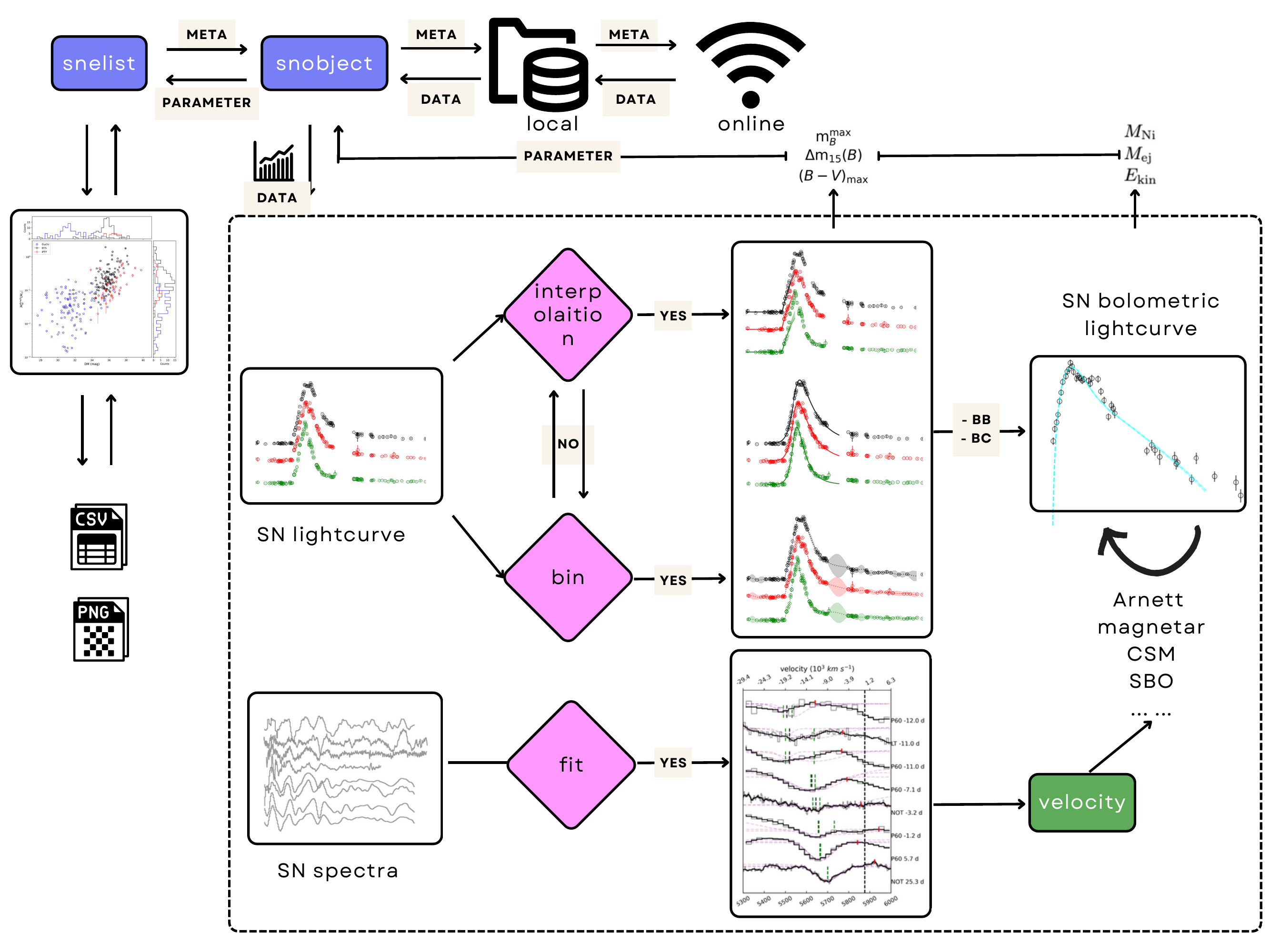}
    \caption{Flowchart of the main steps in the \haffet process, performed on the stripped envelop SNe in Paper I. As shown, after getting observational data for a single SN, {\tt snobject} is called to fit on multi-band LCs, which could be used to estimate photometric properties, as well as provide interpolated magnitudes for estimating color and bolometric LC. A further step is to fit bolometric LC with, e.g. the Arnett model, and spectral lines can be also fitted for photospheric velocities, which are often used to help break the degeneracy of the Arnett model. Finally, all the fitted parameters from model fittings are collected by the {\tt snelist} for a sample study.}
    \label{fig:flowchart}
\end{figure*}

For better clarity, in Figure~\ref{fig:flowchart}, we illustrate the main steps as they were generally performed using \haffet for stripped envelope SNe in Paper I: 

\begin{enumerate}

  \item   \haffet initiates a Python class named {\tt snelist}, incorporating a meta table\footnote{In \haffet, we use pandas \citep{mckinney10a} dataframes for table construction: \url{https://pandas.pydata.org/pandas-docs/stable/reference/api/pandas.DataFrame.html}}, which encompasses a list of objects. Within this meta table, every object's meta information, such as coordinates and redshifts, is standardized, establishing the groundwork for the creation of a subsequent class known as {\tt snobject}. The purpose of the {\tt snobject} class is to manage individual objects.

  \item Within the context of {\tt snobject}, several built-in functions are available to parse both photometric and spectroscopic data, whether they originate from local or remote sources. Depending on the specific objective, users have the flexibility to optionally assign observational data to the object.
    
  \item Observational data is often discrete, influenced by objective factors such as observing conditions and instrument maintenance. As an option, \haffet provides functions that can convert scattered data into a continuous format, thereby aiding in the estimation of specific characteristics, such as the peak magnitude of a lightcurve.
  
  \item Certain photometric features necessitate simultaneous data from multiple distinct wavebands to be acquired. To address this, \haffet incorporates functions (via either binning or interpolations) that enables the alignment of observational data from different wavebands. As an example, color estimation can be achieved through a comparison between two bands, subsequently convertible into luminosities using a bolometric correction function. Additionally, matching more than two bands enables the estimation of Spectral Energy Distribution (SED) and corresponding luminosities.
 
  \item All photometric features, whether derived directly or indirectly, can undergo fitting procedures and subsequent parameter analysis using a range of analytical models.
    
  \item For spectral data, the processing is similar to photometric data, yet comparatively simpler, given their typically one-dimensional nature. Each spectrum is binned, smoothed, and normalized. The continuum can be used to estimate the SED. Simultaneously, the automatic identification of emission and absorption lines takes place, followed by fitting with diverse models to measure spectral attributes, particularly identifying the troughs of absorption features to estimate ejecta velocities.
    
   \item All the photometric and spectral properties from the {\tt snobject} class can be accessed through {\tt snelist}, and subsequently distributed into parameter space for the purpose of sample exploration. Furthermore, fitting procedures can also be executed within this context.
      
\end{enumerate}

In essence, \haffet serves as a versatile tool for transient data analysis, engineered to aid researchers in processing and comprehensively analyzing an array of transient data types. Its modular design and assortment of operating modes render it adaptable to a wide spectrum of research inquiries and analytical requirements. We would like to mention that
the functions and their usages of {\tt snobject} and {\tt snelist} can be found in the online documentation\footnote{For example, \url{https://haffet.readthedocs.io/en/latest/snobject.html} for {\tt snobject}.}. In this paper, we give a theoretical overview of \haffet, and users should consult the corresponding functions for each section on the documentation page for more information and application examples. 

\section{Data Input}
\label{sec:data}

As discussed in Sec. \ref{sec:haffet}, two different kinds of data, i.e. meta and observational data, might be employed in the fitting procedure. 
These data typically originate from various sources, leading to significant differences in their formats. In the meanwhile, users may work with their own data. Hence, in this chapter, our focus lies on the standardized formats for various data types within \haffet. We will demonstrate the normalization procedure to accommodate these variations.

\subsection{Meta data}
\label{sec:meta}

Users may want to supply additional information for the SNe in addition to the observational data, for example spatial coordinates for milky way extinction or the redshifts to correct observing data to the rest frame.
Within \haffet, we have integrated pre-deployed metadata tables sourced from various open-source databases, e.g. 
the Bright Transient Survey explorer\footnote{\url{https://sites.astro.caltech.edu/ztf/bts/explorer.php}} (BTS) catalog and the Open Astronomy Catalog\footnote{\url{https://github.com/astrocatalogs/OACAPI}} (OAC)  tables. Users can also create their own tables specifically for their needs, with proper column names following this specific scheme\footnote{More keywords of the scheme can be found at the {\tt snelist} part of \href{https://github.com/saberyoung/HAFFET/blob/master/sdapy/data/default_par.txt}{https://github.com/saberyoung/HAFFET/blob/master/sdapy/data/default\_par.txt}.}:

\begin{verbatim}
... ...
idkey  : objid # str  # meta key as index
idkey1 : alias # str  # meta key as alias
rakey  : ra    # str  # meta key for ra
deckey : dec   # str  # meta key for dec
zkey   : z     # str  # meta key for redshift
distkey: dist  # str  # meta key for distance
... ...
\end{verbatim}

As shown, we compile the key, value, type, and some annotations of the data tables into a readable text.
Below is an illustration showing how only columns with certain names can be appropriately identified and processed 
to the \haffet meta table.
Here, the object ID and coordinates would be read correctly, but the redshifts would be incorrect because of the erroneous column name (redshift instead of $z$).

\begin{verbatim}
objid,ra,dec,redshift
ZTF18aajpjdi,13:18:06.62,+27:52:24.2,0.0746
ZTF18aahtkze,11:14:19.24,+34:28:43.7,0.0693
ZTF18aapictz,11:59:37.00,+27:31:50.0,0.084
... ...
\end{verbatim}

The Object ID is the only item 
required as the identifier of an object. 
In most cases, an object would have multiple names, e.g. the survey name and the IAU name. 
In order to compile all of these names that may be used to search for a certain object, an alias ID column is established as an option. 

Coordinates are essential mainly when using forced photometry services, and for getting Milky Way extinctions. There are built-in functions that can handle coordinates with different formats via astropy \citep{astropy:2013, astropy:2018, astropy:2022} \footnote{\url{https://docs.astropy.org/en/stable/index.html}}, as well as the healpix \citep{healpix} \footnote{\url{https://healpy.readthedocs.io/en/latest/index.html}} indices for mollweide plots.  

The redshift for each object should be well determined; otherwise all the 
analysis will be done in the observer frame.
In the framework of \haffet, users have the flexibility to define cosmological assumptions, parameters, error budgets, etc, in order to determine distances based on input redshift information. Here, we exemplify this by showcasing the default \haffet settings, that is also the specific configuration applied to stripped envelope SNe in Paper I.
Unless distances are specified, redshifts were converted to distances using a flat cosmology with H$_0=70$~km~s$^{-1}$~Mpc$^{-1}$ and $\Omega_{\rm{m}} = 0.3$. This was done using the redshifts in the CMB frame where the dipole parameters were obtained from \cite{Fixsen}. We also included corrections for peculiar velocities using the formalism of \cite{2015MNRAS.450..317C}.
For each individual SN we include an uncertainty on the peak magnitude corresponding to a 150~km~s$^{-1}$ uncertainty included in the peculiar velocity correction and a systematic $\pm3$~km~s$^{-1}$~Mpc$^{-1}$ error on the adopted Hubble constant. 
For the most nearby objects, with the largest peculiar motion corrections, the relative errors on the absolute magnitudes are the largest.
Comparing the estimates for peculiar velocity corrections for our SNe (Paper I) within $z < 0.015$ with values estimated using \cite{2015ApJ...807...35M} and the corrections provided by NED for the Virgo, Great attractor and Shapley cluster \citep{Mould2000ApJ...529..786M} we find a rms of 0.19 mag, so for these low-z SNe we add another 0.15 mag to the peak luminosity uncertainty in our error budget.

The transient type matters for some of the analysis, for example which spectral line should be used to determine the photospheric velocity, or which intrinsic colour template should be compared with to estimate the host galaxy reddening. 
\haffet is able to supply this from the BTS or OAC catalogs.
There is also a straightforward photometric categorization procedure within \haffet as implemented from 
astrorapid \citep{astrorapid} \footnote{\url{https://astrorapid.readthedocs.io/en/latest/index.html}}. Moreover, since \haffet provides spectral line identification, this is aiding in a more secure classification of the transient types.

By default, \haffet would ignore extinctions unless they are explicitly assigned or inferred using \haffet functions.
\haffet correct all photometry for Galactic extinction, using the Milky Way (MW) color excess $E(B-V)_{\mathrm{MW}}$ toward the position of the SNe, unless precise coordinates have been assigned.
These are all obtained from \cite{2011ApJ...737..103S}. All reddening corrections are applied using the \cite{1989ApJ...345..245C} extinction law with $R_V=3.1$. 
\haffet utilizes dustmap\citep{dustmaps} package for Galactic extinction estimation, and users can refer to their documentation page\footnote{\url{https://dustmaps.readthedocs.io/en/latest/}} for a wider selection of maps and parameter configurations.
Host galaxy extinction is more uncertain. One method to estimate host-galaxy extinction is to make use of the fact that SNe often have similar intrinsic colors after peak \citep[e.g.,][]{Drout2011}. This was developed by \cite{stritzingercolors} and \cite{taddia2018csp} for stripped envelope SNe and Type II SNe. 
For super-luminous SNe (SLSNe), \citet[][their table A4]{zhihao} investigated the $g-r$ colours at peak for a large sample of ZTF objects. 
To estimate the values for the host-galaxy extinction, \haffet collected  these intrinsic SN color-curve templates and attribute the difference between the observed and the intrinsic color to the host galaxy reddening.

\subsection{Observational data}
\label{sec:obsdata}

\begin{figure*}
\centering
    \includegraphics[width=0.6\textwidth]{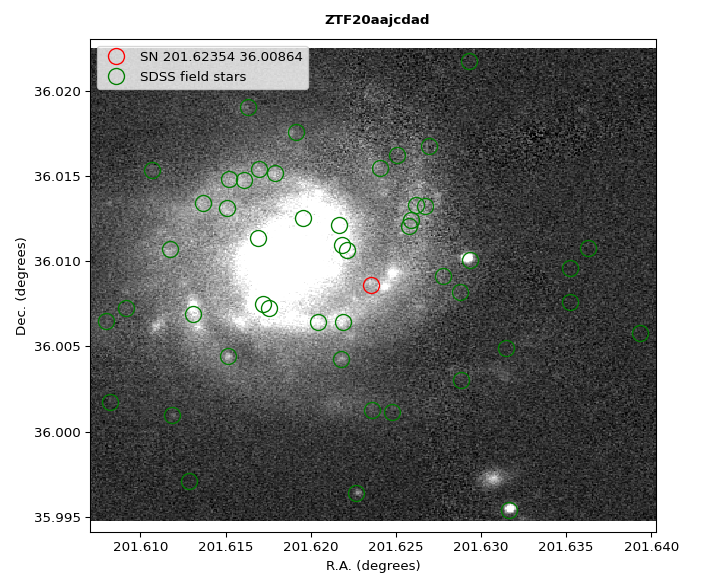}
    \caption{The image of the field of SN 2020bcq as a finder that was automatically made by \haffet. The images are queried with the PS1 Image Cutout Service\tablenotemark{a}, while the SDSS field stars (the green open circles) are downloaded via astroquery\tablenotemark{b}. The position of SN 2020bcq is marked as the red source in the middle of the inset.}
    \label{fig:finder}
\tablenotetext{a}{\url{https://outerspace.stsci.edu/display/PANSTARRS/PS1+Image+Cutout+Service}}
\tablenotetext{b}{\url{https://astroquery.readthedocs.io/en/latest/}}
\end{figure*}

Once a list of objects is constructed using an appropriate meta table, each object's observational data must be assigned. 
Besides the image (see Fig. \ref{fig:finder}),
\haffet can be utilized to work with photometric and spectroscopic data, and in this section we describe how to input these data to \haffet.

\subsubsection{Using private data and arbitrary input formats}
\label{sec:private}

Different datasets may have significantly different data formats, depending on how the data was reduced and presented. 
For example, certain codes favor the \texttt{json} format, while others make use of the \texttt{csv} format.
Many sources may also use various keywords to describe the data, e.g. different notations of magnitudes or fluxes with distinct units.

By asking for particular previously determined keywords, \haffet uses a pandas dataframe to cope with pure text forms and streamline the procedure.
In the example shown below, the photometric data file should use the same keywords in order for them to be properly read\footnote{More keywords of the scheme can be found at the {\tt snobject} part of \href{https://github.com/saberyoung/HAFFET/blob/master/sdapy/data/default_par.txt}{https://github.com/saberyoung/HAFFET/blob/master/sdapy/data/default\_par.txt}.}:

\begin{verbatim}
... ...
jdkey    : jdobs # str # julian date
magkey   : mag   # str # AB magnitude
emagkey  : emag  # str # magnitude error
limmagkey: limmag# str # limiting magnitude
fluxkey  : flux  # str # flux (micro Jy)
efluxkey : eflux # str # flux error
filterkey: filter# str # filters
... ...
\end{verbatim}

It is important to note that \texttt{jdkey} and \texttt{filterkey} are strictly required, and either mag/emag or flux/eflux should be provided after that.
Then, to convert between magnitudes and fluxes, built-in functions can be utilized. 
\haffet uses a default zeropoint of 23.9, which is appropriate for converting fluxes measured in micro Janskys.

It frequently happens that LCs from different sources do not match up. This can be due to a variety of factors, such as the methodology used to estimate the photometry, the way standard stars were chosen, or not correcting for the different filter functions. Therefore, \haffet makes a \texttt{source} column whenever a LC is appropriately included in order to display the data origins.

Similar plain text files with either 2 or 3 columns are required for input of spectra, as well as an observation time that might be used later to estimate the line velocity evolution. 
The three columns present the wavelengths, fluxes, and their errors. When making spectral fits, \haffet has built-in methods to mimic flux errors depending on the spectral quality if errors are not provided.
The flux units for spectra could be arbitrary, and \haffet is able to absolute calibrate them with interpolated photometric magnitudes.

\subsubsection{Using data from public online sources}
\label{sec:gm}

Since open web servers offer data in a consistent data format, \haffet contains built-in functions for obtaining data from them, e.g.  the ZTF \citep{yuhan2019} and Asteroid Terrestrial-impact Last Alert System (ATLAS) \citep{atlasfp} Forced Photometry Services, the application programming interface (API) of Weizmann Interactive Supernova Data Repository (WISeREP) \citep{Yaron:2012aa}, Open Astronomy Catalog (OAC) \citep{oac} and Transient Name Server (TNS) \citep{tnsfp}, and the Growth marshal \citep{Kasliwal2019}/fritz \footnote{\href{https://fritz.science}{https://fritz.science}} broker for the internal ZTF groups\footnote{Users can easily add sources from their own API in the \haffet package following a consistent data format, and feel free to contact us for adding them into the GUI.}. A valid set of authorizations is required by \haffet to access these services if a user wants to utilize them. 
Guidance on handling these authorizations can be found in the online documentation. \\

\begin{figure*}
\centering
    \includegraphics[width=0.6\textwidth]{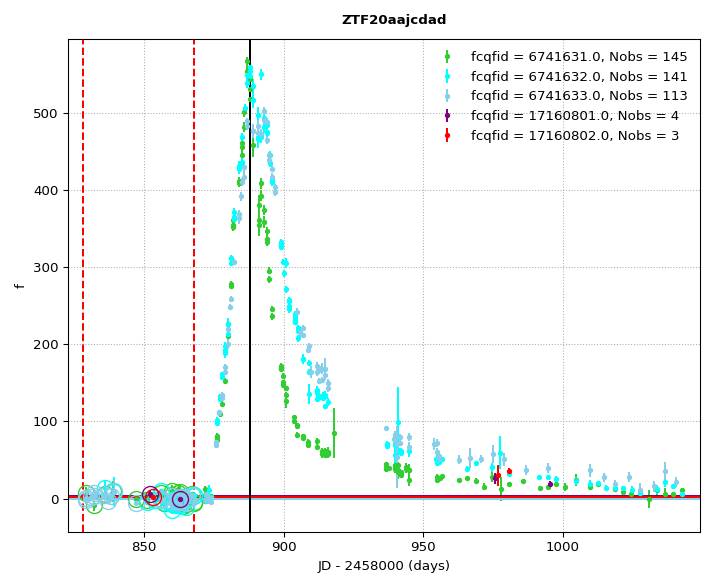}
    \caption{Baseline correction for ZTF forced photometry LCs exemplified by SN 2020bcq. This target was observed in three filters (g, r and i band), and two fields (field 674 and 171). Observations associated with different fcqf ID (using Eq. 1 of \citealp{yuhan2019}) and counts are shown in distinctive colors. The two red vertical dashed lines are set based on the user option for the selection of baseline range. The observations within the range are used to set the baseline flux levels, which are aligned by \haffet so that observations from different fields for the same filter are mutually consistent.}
    \label{fig:ztfbase}
\end{figure*}

\haffet has the capability to preprocess these data, e.g. binning, outlier removal, etc, aiming to standardize them to the best extent possible. Meanwhile, \haffet also incorporates nested auxiliary tools that facilitate appropriate optimization of these data. For instance, as
 shown in Fig. \ref{fig:ztfbase}, for the ZTF Forced Photometry Services the user can use \haffet to check if the object is covered by multiple fields, and if so, correct the baseline of different fields in the same filter. More details can be found in \citet[][see their Fig. 4]{yuhan2019}.

\section{Model Fittings}
\label{sec:fitting}

 Once observational data for a SN have been acquired and standardized, the subsequent stage in \haffet involves fitting the data using a range of analytical models.
\haffet is built to fit functions to a variety of datasets, including spectral absorption features and many kinds of LCs.
Despite having very different data, these can be fitted using the same techniques.
In \haffet we therefore specify a number of "models" that provide the objectives for the data to approach as well as so-called "engines" that are used to prepare the dataset for certain fits.

There are currently nine engines available in \haffet, and these are shown in 
Table~\ref{tab:engines}, along with a variety of built-in models for each of them, see Table~\ref{tab:models}.
In this section, we describe all built-in engines and models, show  the situations in which these engines and models should be utilized, 
how to tune their hyperparameters, and explain how users can define their own engines and models.

\begin{table*}
\caption{Table of built-in engines defined in \haffet.}
\begin{center}
\begin{tabular}{cc} \hline\hline
Engine name & Description \\ 
\hline
multiband\_early & multi-band LCs at early phases \\ 
multiband\_main & multi-band LCs at around the peak epoch \\ 
bol\_early & bolometric LCs at early phases \\ 
bol\_main & bolometric LCs at around the peak epoch \\ 
bol\_tail & bolometric LCs at the tail phases \\ 
bol\_full & bolometric LCs for the full range \\ 
sed & spectral energy distribution data, \\ 
& i.e. multi-band photometry or spectra \\ 
specline & spectral line absorption feature \\ 
specv\_evolution & spectral/photospheric line velocities \\
\hline
\end{tabular}
\label{tab:engines}
\end{center}
\end{table*}

\begin{table*}
\caption{Table of built-in models defined in \haffet. Here we list only the basic models, and in \haffet there are also models with multiple components, e.g. the Arnett model at the main peak plus the tail model, and a boundary of time between them is set as a free parameter as well. For such a case, we use a bol\_full engine to prepare bolometric LCs for model fitting.}
\begin{center}
\begin{tabular}{cccc} \hline\hline
Model Name & Engine & Description & Reference(s) \\ 
\hline
bazin & multiband\_main & parametric model suited for SNe Ia LCs & \cite{bazin} \\ 
villar & multiband\_main & parametric model for a broad range of LC & \cite{villar} \\ 
 & & morphologies, e.g. with bumps or plateau & \\ 
risepl & multiband\_early & power law model for the early rise of SNe & \cite{Miller_2020} \\ 
salt2 & multiband\_main & empirical model for SNe Ia LCs & \cite{salt2} \\
arnett & bol\_main & radioactive model for bolometric LCs & \cite{arnett1982}  \\ 
 &  & of SNe Ia or CC at around the peak epochs & \\ 
tail & bol\_tail & radioactive model for bolometric LCs & \cite{Wygoda2019} \\ 
 & & of SNe Ia or CC at the tail phases & \\ 
sbo & bol\_early & shock cooling tail model for SNe IIb/Ib & \cite{piro2021} \\ 
magnetar & bol\_main & magnetar model for SLSNe & \cite{Nicholl_2017} \\ 
gauss & specline & Gaussian model for spectral line profile & \cite{gauss} \\ 
voigt & specline & Voigt model for spectral line profile  & \cite{voigt} \\ 
exponential & specv\_evolution & exponential model for velocity evolution & \cite{Fremling2018} \\
\hline
\end{tabular}
\label{tab:models}
\end{center}
\end{table*}

\begin{figure*}
\centering
    \includegraphics[width=0.6\textwidth]{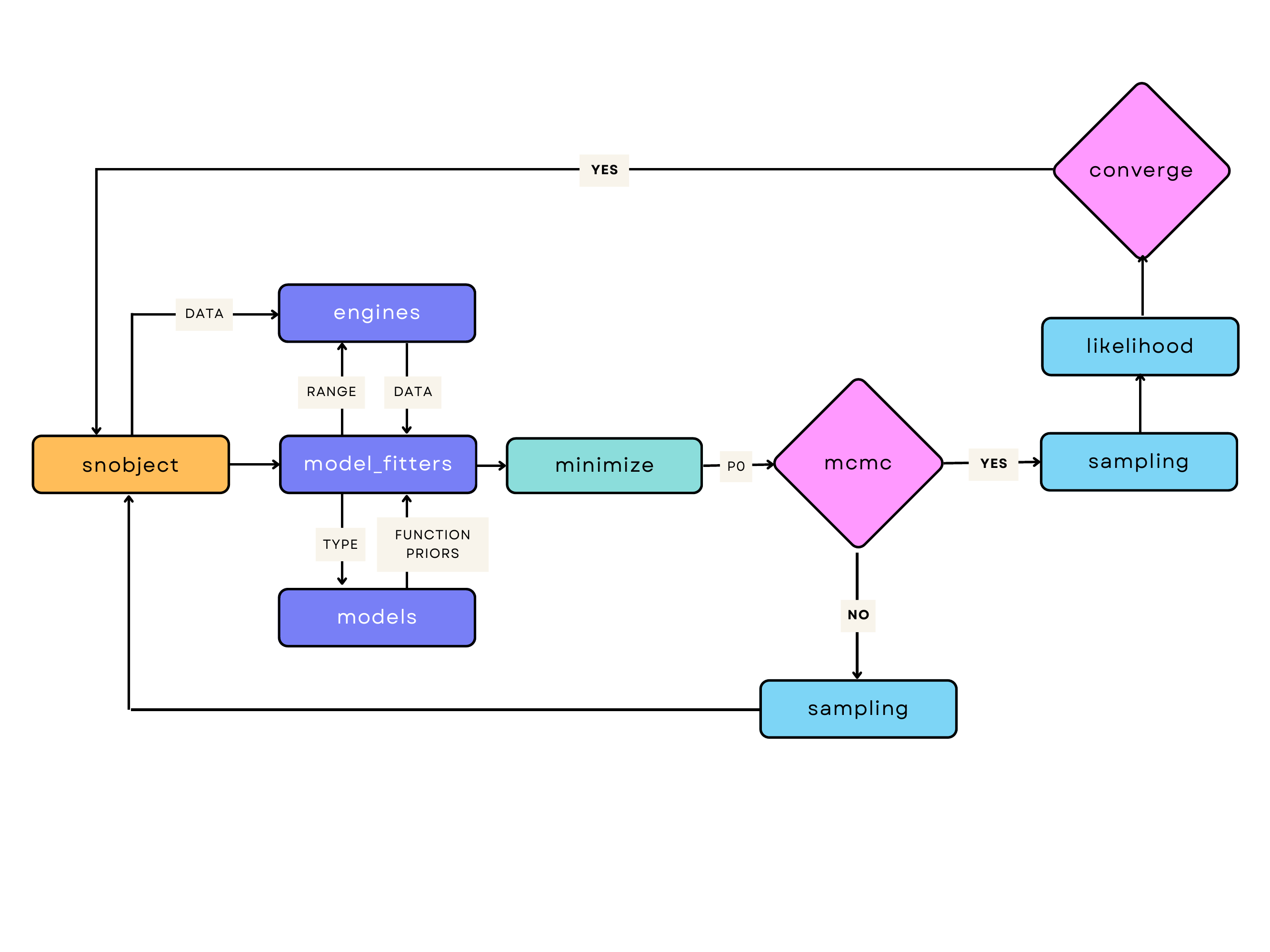}
    \caption{Flowchart illustrating model fitting procedures of \haffet. As shown, \haffet first obtain model function and priors, as well as data to be fitted that were prepared by the engines. Then a minimizing process is performed to constrain model parameters. The output could be used as prior for MCMC routines. Finally the fitted samplings can be called by \haffet to, e.g. produce parameter contours, and reproduce observational data.}
    \label{fig:modelfit}
\end{figure*}

\subsection{Defining models}
\label{sec:model}

A \texttt{python} file that contains analytical functions and a \texttt{json} file that specifies the engine and parameters are used to define each model in \haffet.
Below is an example using the \texttt{json} file of the power-law fits (which is further discussed in section~\ref{sec:multiearly}):

\begin{verbatim}

modelmeta = {
    ...
    'powerlaw_multiple':
    {
        'engine': 'multiband_early',
        'alias': ['pl', 'powerlaw',],
        'description': 'Power law fits',
        'func': 'functions.powerlaw_full',       
        'parname' :[ 
            r'$t_\mathrm{fl}$', 
            r"$A$",  
            r"$\alpha$", 
            r"$C$",],
        'par' :['texp', 'a', 'alpha', 'c',],
        'bestv':{
            'texp' : -18,
            'a' : 100,
            'alpha' : 2,
            'c' : 60,},
        'bounds': {
            'texp' : [-100, 0],
            'a' : [0, 10000],
            'alpha' : [0, 10],
            'c' : [0, 10000],},
        'same_parameter': ['texp'],
        'fit_error' : True,
    },
    ...
}
\end{verbatim}

As shown, the \texttt{powerlaw\_multiple} model would be called to fit data prepared by the \texttt{multiband\_early} engine with the \texttt{powerlaw\_full} functions having four free parameters. As discussed in Section~\ref{sec:optimize}, there are two different likelihoods available in \haffet. By default, \haffet uses a Gaussian likelihood, while if we set \texttt{fit\_error} = True, it will include one more term to the likelihood as additional model uncertainties. 
The \texttt{same\_parameter} item decides whether to fit multiple band LCs simultaneously or independently.  If the \texttt{same\_parameter} is left blank, \haffet will fit the LCs in different bands one at a time. If some other parameters are specified, such as \texttt{t\_exp} for example, \haffet fit all bands at once with the same \texttt{t\_exp}, which is the epoch of first light for such a case.
 We have showcased examples of how to define engines and models on the documentation page, which users can always refer to.
  
\subsection{Built-in models}
\label{sec:builtin}

In the sections that follow we outline the fundamental concepts of the built-in models while highlighting the possibility of experimenting with various fitting choices, such as different free parameters, likelihoods, or fitting patterns. 

\subsubsection{multiband\_early models}
\label{sec:multiearly}

'multiband\_early' engine will extract multi-band light curves of SNe, automatically identify the baselines, and, based on user settings, extract early-phase multi-band light curves from baseline to a specified percentage of the estimated peak.
As a representative model,
power-law fits are working with multiple-band photometric data prepared by the 'multiband\_early' engine. 
These fits can for example be used to estimate explosion epochs and rise times, but the precision by which we can constrain the first light is also of importance for many other investigations (for example for coincidence studies with neutrino signals or gravitational wave detections).
\cite{Miller_2020} developed a Bayesian framework to model the early rise of LCs as a power law in time in their Eq. 1:
\begin{equation}
    f_b(t) = C + H[t_\mathrm{fl}] A_b (t - t_\mathrm{fl})^{\alpha_b},
    \label{eqn:flux_model}
\end{equation}
where C is the baseline flux present
in the reference image prior to SN discovery, $t_\mathrm{fl}$ is the time of first light, $H[t_\mathrm{fl}]$ is the Heaviside function equal to 0 for t $< t_\mathrm{fl}$ and 1 for t  $\geq  t_\mathrm{fl}$, $A_\mathrm{b}$ is a constant of proportionality in filter b, and $\alpha_\mathrm{b}$ is the power-law index in filter b. 

\cite{Miller_2020} applied this framework to a subset of normal SNe Ia from ZTF to simultaneoulsy model the evolution in both the ZTF $g$ and $r$ filters. 
This assumes that the epoch of first light is  
the same in the two ZTF bands, which is a fair assumption given the ZTF cadence and the similarity of the SN ejecta opacity at these wavelengths. 
In \haffet we follow the same methodology to characterize the early emission in the \texttt{powerlaw\_multiple} model (see Sect. \ref{sec:model}), but also provide the possibility to fit the different band LCs independently by removing \texttt{same\_parameter} in the \texttt{power\_single} model. 
This configuration is maintained as the default setting in \haffet and is ultimately applicable across all models.

By way of illustration, we employed this model to estimate the explosion epochs of SNe in Paper 1.
As in \cite{Olling2015}, \cite{Miller_2020} only include observations up to 40\% of the peak
amplitude of the SN LC, and note that this particular choice, instead of 30\% or 50\%, does slightly affect the final inference for the model parameters.
For comparison we adopt the same settings as default in the \texttt{multiband\_early} engine, where the peak amplitude in each band should be previously estimated via the gaussian process or other \texttt{multiband\_main} models. 
It is noteworthy that the explosion epoch can also be fitted to bolometric models as a free parameter, or estimated as the middle of the last non-detection and the first detection epochs.
In \haffet, we provide users the option to choose their method of estimation.

\subsubsection{multiband\_main models}
\label{sec:multimain}

SNe LCs typically exhibit a rapid rise to maximum after the explosion, followed by a decline, and eventually accompanied by a linear tail. In the meanwhile unique physical mechanisms, such as CSM (circumstellar material), can result in multiple peaks in the LCs. Currently, there are numerous analytical physical models that can effectively reproduce the evolution of the primary peak in SNe LCs through parameterization.
By fitting these models to the LCs near the main peak, we not only achieve data continuity but also directly describe transient LCs through the parameters of these models.
Within \haffet,  'multiband\_main' engine  extracts data encompassing the main peak of the LC in accordance with user-defined criteria for model fitting. Identifying the primary peak often necessitates techniques that are dissociated from the model itself. By default, \haffet employs the Gaussian regression algorithm (GP) for peak determination. GP is a non-parametric data-driven interpolation technique utilized for serializing LCs.
In order to include color effects, we perform time-series forecasting on the joint multi-band (e.g. $g$ and $r$) fluxes convolved with the wavelengths. 
We use a flat mean function\footnote{Other mean functions are also available in \haffet, such as the Bazin or Villar models.} for the flux form, and establish two kernels, i.e. a constant kernel for the wavelengths, and a stationary Matern32 kernel for the fluxes, with the {\tt GEORGE}\footnote{\href{https://george.readthedocs.io}{https://george.readthedocs.io}} package  \citep{george}.

The GP interpolation becomes less informative with larger uncertainties and when the data are sparser. For estimating physical parameters for normal SNe, an alternative approach is then to fit the data using analytic functions, such as those used by \cite{bazin, Karpenka, kessler1, kessler2, villar}.
For this purpose we included the \cite{bazin} and \cite{villar} models in \haffet. The 'Bazin' model is appropriate for SN LCs without fluctuations, whereas the 'Villar' model is also useful to fit plateaus and bumps.
For the Bazin model, the flux form is determined by 5 parameters:

\begin{equation}
\label{eqn:bazin}
F(t) = \mathrm{A} \frac{e^{-(t-t_0)/\tau_\mathrm{fall}}}{1+e^{-(t-t_0)/\tau_\mathrm{rise}}} + \mathrm{B}
\end{equation}
where \citet[][their fig.~8]{taddia2015sdss} illustrate the different parameters and how they contribute to the light curve properties.
We can obtain the date of maximum from the derivative of this function as $t_\mathrm{max} = t_0 + \tau_\mathrm{rise} \mathrm{log} \frac{-\tau_\mathrm{rise}}{\tau_\mathrm{rise}+\tau_\mathrm{fall}}$ \citep{bazin}.

The functional form of the Villar model is similar to the Bazin model, but incorporates a plateau component by adding two additional free parameters:

\begin{equation}
\label{eqn:villar}
F(t) = \begin{cases} 
      \frac{A+\beta(t-t_0)}{1+e^{-(t-t_0)/\tau_\mathrm{rise}}} & t<t_1 \\
      \frac{\left(A+\beta\left(t_1-t_0\right)\right)e^{-(t-t_1)/\tau_\mathrm{fall}}}{1+e^{-(t-t_0)/\tau_\mathrm{rise}}} & t\ge t_1 
   \end{cases}
\end{equation}

The comparison of the Villar model with the Bazin model and the \cite{Karpenka} model are shown in \citet[][their fig.~3]{villar}. 

\subsubsection{SED models}
\label{sec:lum}

The SED engine is designed to match multi-band data in the temporal dimension, resulting in wavelength-dependent data at different epochs. This allows for color calculation, blackbody fitting, etc, which can subsequently be utilized to estimate bolometric luminosities. Taking a different perspective: 
three approaches are adopted in \haffet to estimate the bolometric luminosity based on the extinction-corrected brightness and the distance. 

First, we use the bolometric correction (BC) coefficients of \cite{Lyman14, lyman16}. These color-dependent coefficients were measured by fitting the spectral energy distributions of a sample of stripped-envelope SNe that have good observational coverage at ultra-violet, optical, and infrared wavelengths, and can be utilized as long as the light curves for a SN of interest are available in at least two bands. The bolometric magnitude is expressed as
\begin{equation}
  M_{bol} = M_{\odot,bol}-2.5\log_{10}\left(\frac{L_{bol}}{L_{\odot,bol}}\right),
\label{eq:bolo}
\end{equation}
where $L_{bol}$ is the bolometric luminosity, $L_{\odot,bol}$ and $M_{\odot,bol}$ are the luminosity and bolometric magnitude of the Sun.
A BC to filter $x$ is defined as:
\begin{equation}
  \text{BC}_{x} = M_{bol} - M_{x},
\label{eq:bc}
\end{equation}
where $M_{x}$ is the absolute magnitude of the SN in filter \emph{x}. 
For example, if a  SE SN is monitored in the $g$ and $r$ filters, then according to \citet[][their table 2]{Lyman14}, this SE SN during the photospheric phases ($-0.3~{\rm mag} < g-r < 1.0~{\rm mag}$) will have
\begin{equation}
  \text{BC}_{g} = 0.054 - 0.195 \times (g-r) - 0.719 \times (g-r)^{2}
  \label{eq:bc_se_sl}
\end{equation}
In this way, the bolometric LCs of the SNe can be estimated by matching up the $g$ and $r$-band magnitudes.

Three approaches in \haffet are utilized for the color epoch match-up, i.e. binning, GP, and fitting. \haffet projects $g$- and $r$-band detections onto a grid (with a user-defined binning, such as 1 day) that may be readily matched.
The binning method is taking all epochs with both bands available, and convert them to bolometric magnitudes and hence luminosities.
For epochs with only one band, \haffet can estimate magnitudes for the missing band from either GP interpolation or from analytic fittings.

The above match-ups can also be applied to multiple filters. 
In \haffet, we can construct SEDs as a diluted blackbody (BB) distribution for those epochs with multiple-band data, and the luminosity can then be calculated by integrating the fitted BB from 2000 \ang to 20000 \ang\footnote{These are the defaults, but can easily be revised in the configuration file.}.
We fit a diluted BB function with multiple bands using the following formula\footnote{The filter information, e.g. the central wavelengths and widths, are taken from the superbol \citep{superbol} package.}:
\begin{equation}
    F_{\lambda}=(R/d)^2 \cdot \epsilon^2 \cdot \pi \cdot B(\lambda, T) \times 10^{-0.4 \cdot A_{\lambda}},
\end{equation}
where $F_{\lambda}$ is the flux at wavelength $\lambda$, $B$ is the Planck function, $A_{\lambda}$ is the extinction in magnitudes, $T$ is the temperature, $R$ is the radius, $d$ is the distance, and
$\epsilon$ is the dilution factor \citep{E96, Hamuy01, D05} that represents a general correction between the fitted BB distribution to the observed fluxes. We use the values from \citet{D05} in \haffet for SNe II. For other types of SNe, the default is to ignore the dilution factor but users can also use their own functions.

Finally, the derived magnitudes (from section \ref{sec:multiearly} and \ref{sec:multimain}) can also be used to absolute calibrate the input spectra, which could then be used to fit the SED models (e.g. the blackbody function) or integrated for bolometric luminosities. For this approach, we have included the transmission information for ZTF $g$, $r$ and $i$ filters in \haffet.
After obtaining the bolometric LCs, there are a variety of models available in \haffet that can fit them.

\subsubsection{bol\_early models}
\label{sec:sbo}

Following shock breakout, the emission from an astrophysical explosion can be dominated by the radiation of shock-heated material as it expands and cools, known as shock cooling emission. 
This is often  seen for a subgroup of supernovae, SNe IIb, that exhibits double-peaked light curves, but there is also evidence for this kind of early emission from other transients, presumably with extended atmospheres.

In \haffet, similar to the multiband\_early engine, bol\_early engine primarily focuses on extracting the early-phase bolometric LCs that was inferred from the SED engine. As a default built-in model,
we implement the shock cooling emission model from \cite{piro2021}, that fits LCs prepared by the 'bol\_early' engine, and here the luminosity can be expressed as \citep[][their functions 17 and 20]{piro2021}:

\begin{equation}
\label{eqn:sbo}
L(t) = \begin{cases} 
      {\rm P}~(\frac{t_d}{t})^{4/(n-2)} & t<t_d \\
      {\rm P}~exp\left[ (-0.5~(\frac{t}{t_d})^2-1) \right] & t\ge t_d 
   \end{cases}
\end{equation}

where the prefactor $P = \frac{\pi(n-1)}{3(n-5)} \frac{c~R_e~v_t^2}{\kappa}$, 

$t_d = \left[ \frac{3~\kappa~K~M_e}{(n-1)v_tc} \right]^{1/2}$, 

and $v_t = \left[ \frac{(n-5)(5-\delta)}{(n-3)(3-\delta)} \right]^{1/2}~(\frac{2E_e}{M_e})^{1/2}$. 
Here, $n=10$, $\delta=1.1$, $K=0.119$, and $t$ is the time relative to the explosion epoch.

Due to the hot temperatures during shock cooling emission, the function can be simplified by assuming a constant electron scattering opacity $\kappa$.
\cite{piro2021} use $\kappa=0.2~{\rm cm^2g^{-1}}$ for their helium-rich SNe.
In \haffet, we could either fit the opacity as a free parameter, or use a constant opacity following \citet[][their table 6]{Nagy2018}:
$\kappa=0.21~{\rm cm^2g^{-1}}$ for SNe IIP, $0.20~{\rm cm^2g^{-1}}$ for IIb, $0.18~{\rm cm^2g^{-1}}$ for Type Ib, and $0.10~{\rm cm^2g^{-1}}$ for Type Ic SNe, or use constant opacity from user input, e.g. for the ZTF Ibc sample study in Paper I, we use a constant opacity $\kappa=0.07~{\rm cm^2g^{-1}}$ for both SNe Ib and Ic following \citet{Chugai2000,leo2021,barbarino} \footnote{Examining potential variations between the SN Ib and Ic progenitors is one of the objectives of Paper I. Therefore, it is critical to make be able to track if the outcome is dependent of the assumptions, e.g., in assuming the same opacity for both SNe Ib and Ic, but also if the proceedures for host extinction corrections or velocity estimates differ.}.

\subsubsection{bol\_main models}
\label{sec:arnett}

The bolometric LCs of SNe contain a rich array of physical information, and the mathematical modeling of these curves has a well-established history since  \cite{arnett1982}. Within \haffet, we established 'bol\_' engines, that was similar to 'multiband\_' engines, to prepare bolometric LCs in different phases for model fittings.
As default of \haffet, we provide built-in functions to compare the constructed bolometric LCs with the semi-analytic models from \cite{arnett1982} for normal core-collapse SNe for epochs around the main peak.

For a normal SN after the explosion, as the ejecta expand, the decay of \Ni~ to \Co~ and afterward to \Fe~ releases energy into the ejecta which powers the optical LC, including the luminous peak and the later epochs\footnote{We refer to it as the tail phases in this work. When the SN LC reaches a later phase of linear magnitude fall with time, it is said to have reached its LC tail.}. Thus, exploring the bolometric LCs during the peak and tail phases can reveal the properties of the explosion and the progenitor.

With several simplifying assumptions, e.g. the ejecta are expanding homologously, the radiation pressure dominates over the gas pressure, and with a constant opacity and small initial radius, \citet[][their Eq. 36]{arnett1982} provides an expression for all Type I SN LCs, that can be used to estimate the bolometric luminosity with the radioactive inputs, see also \cite{Valenti2008}. 
We assume full $\gamma$-ray trapping since the Arnett formalism is constructed for the LCs around the peak, and get \citep[see][]{chatzo2012, leo2021}: 
\begin{eqnarray}
\label{eq:boloArnett}
L_{ph}(t)=M_{\rm{Ni}}\biggl[(\epsilon_{\rm{Ni}}-\omega\epsilon_{\rm{Co}})\,\Lambda(x,y)+\omega\epsilon_{\rm{Co}}\,\Lambda(x,s)\biggr]
\end{eqnarray}
where 
$x\equiv t/\tau_m$, 
$y\equiv 0.5\tau_m/\tau_{\rm{Ni}}$, 
$s\equiv 0.5\tau_m/\tau_{\rm{Co}}$, 
$\omega\equiv \tau_m/(\tau_{\rm{Co}}-\tau_{\rm{Ni}})$.
Here, $\epsilon_{\rm Ni}= 3.9 \times10^{10}$~erg~g$^{-1}$~s$^{-1}$ and $\epsilon_{\rm Co}=6.8 \times 10^9$~erg~g$^{-1}$~s$^{-1}$ are the specific heating rates of Ni and Co decay, respectively, and $\tau_{\rm{Ni}}=8.8$~days and $\tau_{\rm{Co}}=111.3$~days are their corresponding decay (e-folding) timescales. Using \citet[][his Eq. 31]{arnett1982}:
\begin{eqnarray}
\label{eq:boloArnett_lambda}
\Lambda(x,y)=e^{-x^2}\mathop{{\int}}^x_0e^{-2\rm{z}\rm{y}+\rm{z}^2} 2\rm{z} dz
\end{eqnarray}
$\tau_m$ is the characteristic time scale, and by assuming a uniform density within the ejecta:
\begin{eqnarray}
\label{eq:taum}
\tau_m^2 = \frac{\kappa_{opt}}{\beta c} \sqrt{\Lambda\frac{M_{ej}^3}{E_{kin}}}
\end{eqnarray}
where 
$\beta=13.8$ \citep{Valenti2008}, 
$\Lambda=6/5$ \citep{lyman16}, and
the opacity $\kappa_{opt}$ is discussed at the end of Section 4.2.4.

In \haffet, users have the option to select the free parameters that fit the Arnett model. For instance, the bolometric LCs can be fitted with \Mni + $\tau_m$, or \Mni + $M_{ej}$ + $E_{kin}$, and it is possible to vary the explosion epoch and opacity as free parameters as well. 
The default \haffet model is the \Mni + $\tau_m$, and by doing this, an additional function is needed to break the degeneracy between the ejecta mass ($M_{ej}$) and the kinetic energy ($E_{kin}$) in Eq. \ref{eq:taum}, and for this an estimate for the velocity is employed. 

As discussed in \citet[][their section 5.3]{Dessart2016}, a 'representative' expansion rate, $V_m$ is defined as, 
\begin{eqnarray}
\label{eq:vm}
 V_m\equiv \sqrt{2E_{kin}/M_{ej}}
\end{eqnarray}
that can be used to break the degeneracy between $M_{ej}$ and $E_{kin}$ in Eq. \ref{eq:taum}.
With a large grid of 1D radiative-transfer simulations for SE SNe, they derived analytic expressions between $V_m$ and $V_{ph}$ at the peak epoch as follows:
\begin{eqnarray}
 V_{\rm{ph}}(\rm{He~I~\lambda 5876}) = 2.64 + 0.765~\rm{V_m}, \label{eq:vm1} \\
 V_{\rm{ph}}(\rm{O~I~\lambda7772}) = 2.99 + 0.443~\rm{V_m} \label{eq:vm2}
\end{eqnarray}
and suggest to measure $V_{\rm{ph}}$ from \HeI in SNe IIb/Ib and from \OI in SNe Ic. We derive spectral line velocities ($V_{\rm{ph}}$) in section~\ref{sec:vel}, and follow their methodology to convert $V_{\rm{ph}}$ into $V_{\rm{m}}$ with Eqs.~\ref{eq:vm1} and \ref{eq:vm2}. 
Finally with Eqs. \ref{eq:taum} and \ref{eq:vm}, we can estimate $M_{ej}$ and $E_{kin}$ separately with $V_m$ and the Arnett fitted $\tau_{m}$, as:
\begin{eqnarray}
 M_{\rm{ej}} = \frac{V_m~\tau_m^2~\beta~c}{\kappa_{opt}~\sqrt{2~\Lambda}}, \label{eq:mej} \\
 E_{\rm{kin}} = \frac{V_m^3~\tau_m^2~\beta~c}{\kappa_{opt}~\sqrt{8~\Lambda}}. \label{eq:ek} 
\end{eqnarray}

\begin{figure*}
\centering
    \includegraphics[width=0.45\textwidth]{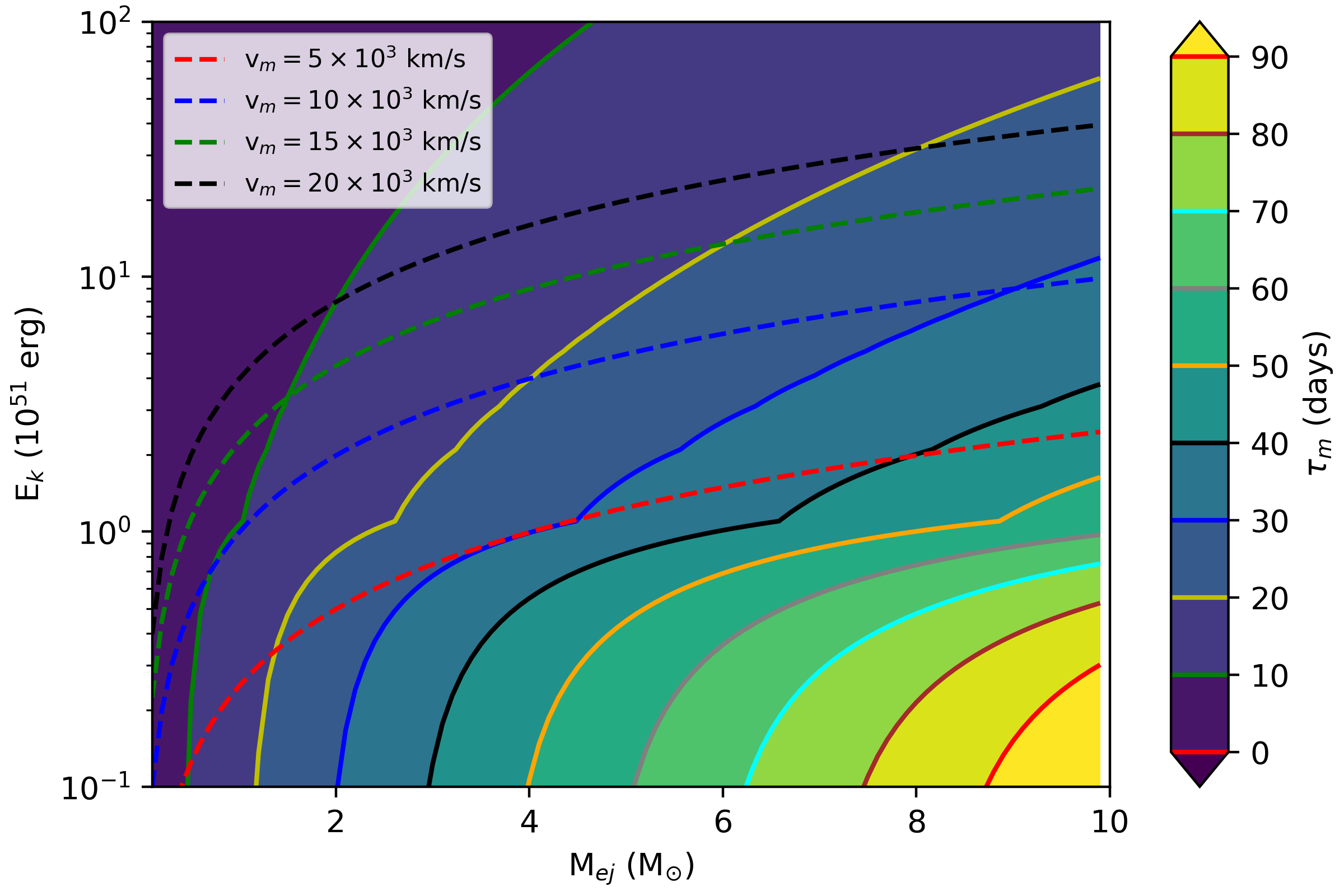}
    \includegraphics[width=0.45\textwidth]{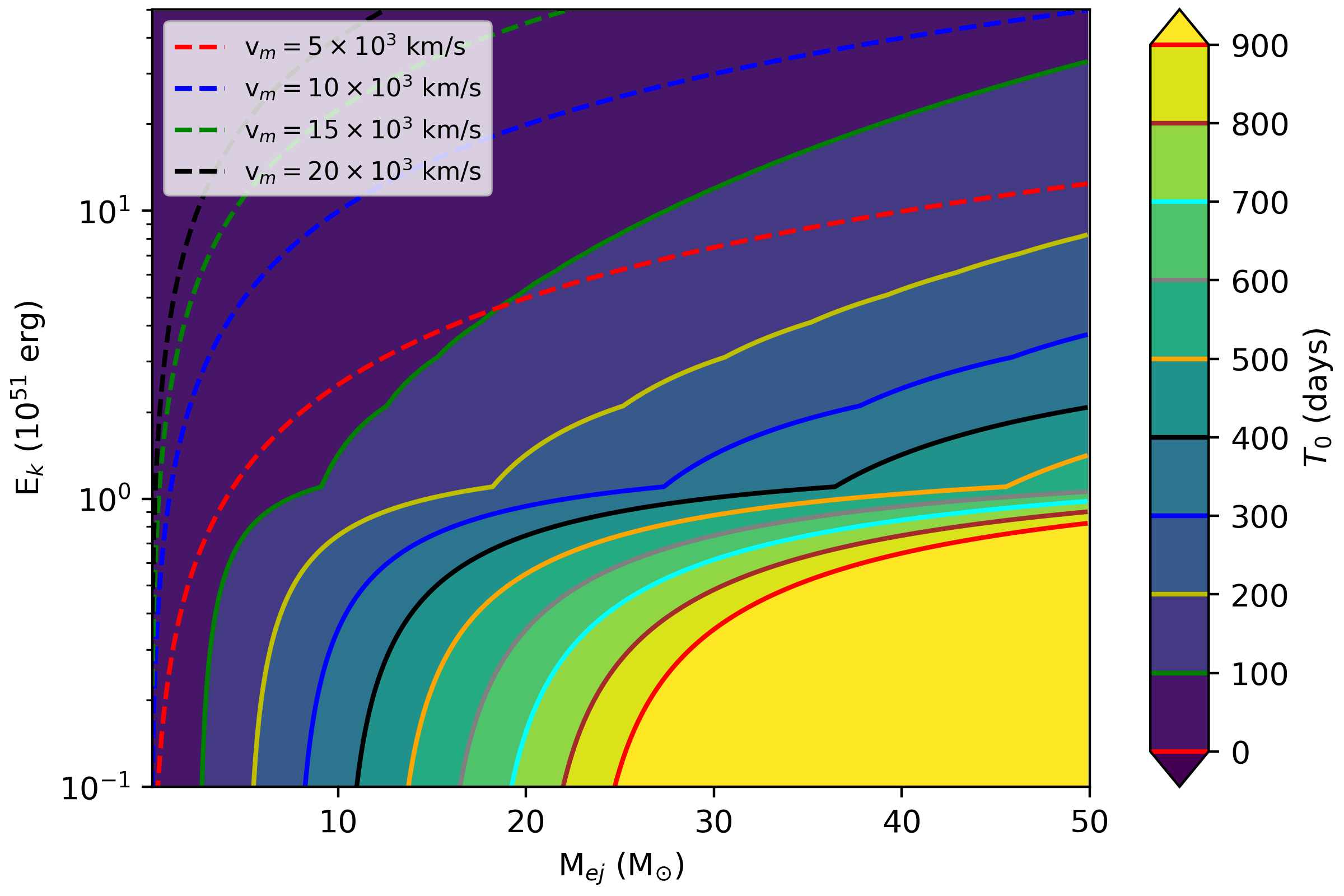}
    \caption{In the left panel, we visualize Eq. \ref{eq:taum} and Eq. \ref{eq:vm}, showing how both $M_{ej}$ and $E_{kin}$ can be estimated with $\tau_m$ together with a velocity measurement. In the right panel, we instead show $T_0$ of the bolometric tail in Eq. \ref{eq:t0}. Here we assume constant opacities, i.e. $\kappa=0.2~{\rm cm^2g^{-1}}$ and $\rm \kappa_\gamma=0.027~cm^2g^{-1}$}
    \label{fig:ekmej}
\end{figure*}

We explore the relationship of $\tau_m$ on $M_{ej}$ and $E_{kin}$ in the left panel of Fig. \ref{fig:ekmej}. As shown, the $M_{ej}$ and $E_{kin}$ can be decided with $\tau_m$ together with a velocity measurement. As example with SN 2020bcq, the fitted $\tau_m$ is $\sim$ 10 days (see Fig. \ref{fig:arnettcompare}) and the peak velocity is  $\sim$ 10,000 $\rm km~s^{-1}$ (see Fig. \ref{fig:velexp}), therefore according to Fig. \ref{fig:ekmej}, $M_{ej}$ and $E_{kin}$ should be around 1 \Msun~ and $1\times 10^{51}$ erg, respectively.

\subsubsection{bol\_tail models}
\label{sec:tail}

An alternative way to estimate the
\Ni~ mass is using the luminosity at the tail, see for example the work by \cite{Afsariardchi2021}.
In \haffet, by default the 'tail' of the LC is defined as the epochs after 60 rest-frame days post-peak.
We fit the tail bolometric LCs with the model from \citet[][their Eqs. 10, 11 and 12]{Wygoda2019}:

\begin{eqnarray}
\label{eq:lni}
&& L \simeq L_\gamma \big(1-e^{-(T_0/t)^2} \big) + L_
 {\rm pos},
\end{eqnarray}
where t is the time since the explosion in the rest frame, $L_\gamma$ and $L_{\rm pos}$ are the total energy release rate of gamma rays and positron kinetic energy respectively.
The term in parenthesis is a deposition factor, which represents the incomplete trapping of gamma rays with $T_0$ as the partial trapping timescale. $T_0$ can be translated to ejecta mass and kinetic energy in a similar way as $\tau_m$ \citep{Clocchiatti1997}:

\begin{eqnarray}
\label{eq:t0}
T_0^2 = \frac{c~\kappa_\gamma~M_{ej}^2}{E_{kin}}
\end{eqnarray}

where $\rm \kappa_\gamma=0.027~cm^2g^{-1}$ is the gamma ray opacity. The luminosity terms can be expressed as:

\begin{eqnarray}
\label{eq:lpos}
&& L_\gamma = M_{\rm Ni} \Big( (\epsilon_{\rm Ni}-\epsilon_{\rm Co})~e^{-t/t_{\rm Ni}}+ \epsilon_{\rm Co}~e^{-t/t_{\rm Co}}\Big), \\
&& L_{\rm pos} = 0.034~M_{\rm Ni}~\epsilon_{\rm Co} \Big( -e^{-t/t_{\rm Ni}} + e^{-t/t_{\rm Co}}\Big)
\end{eqnarray}

In the right panel of Fig.~\ref{fig:ekmej}, we show that the $M_{ej}$ and $E_{kin}$ can be decided with $T_0$ together with a velocity measurement. 

\subsubsection{specline and specv\_evolution models}
\label{sec:vel}

In this section, we describe the specline engine, employed for the preparation of spectroscopic data pertaining to specific elements. Additionally, we detail the procedure for measuring expansion velocities from the data prepared by the specline engine. Furthermore, we discuss the specv\_evolution engine, which facilitates the fitting of velocity evolution to predict velocities at designated phases.

The specline engine in \haffet offers spectral data within a range for model fits. A line's intrinsic wavelength and the permitted velocity ranges can be used by \haffet in addition to a manual input to infer the fitting range.
Such an automated spectral fitting process is shown in the lower left panel of Fig.~\ref{fig:gui}.
After correcting for extinction and time dilation, we fit the continuum of a spectrum with a polynomial, which is afterwards subtracted. We smooth the residual spectrum around the relevant absorption feature with a Savitzky–Golay filter\footnote{Smoothed spectrum is only used for peak search.}, and obtain a number of peak (valley) points via {\tt scipy.signal.find\_peaks} \citep{harris20a} \footnote{\href{https://docs.scipy.org/doc/scipy/reference/generated/scipy.signal.find_peaks.html}{https://docs.scipy.org/doc/scipy/reference/generated/scipy.signal.find\_peaks.html}}. 
The spectra after continuum removal and their local smoothings are shown as the grey and brown lines correspondingly
in Fig.~\ref{fig:gui}.
The vertical dashed line denotes the intrinsic location of a spectral line (Here \HeI for a SN Ib), and the detected peaks and valleys in the spectral region are shown as orange upper and cyan lower arrows.
The peak closest to the intrinsic line location is selected as the red endpoint of the absorption feature. The peak with a slightly lower wavelength is assumed to be the blue boundary of the spectral line. 

\begin{figure*}
\centering
    \includegraphics[width=0.6\textwidth]{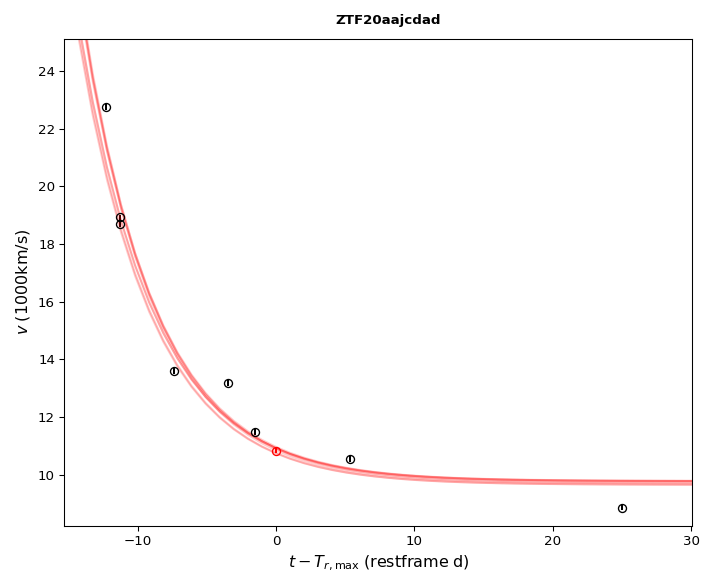}
    \caption{An exponential fit of the fitted velocities on the phases for SN 2020bcq. The red dashed lines are the random samplings of the MCMC, and the red open circle is the estimated velocity at the peak phase from the exponential fit.}
    \label{fig:velexp}
\end{figure*}

The selected fitting range is afterward applied to cut the spectra. After fitting the spectral data within the wavelength range with analytic models, e.g. Gaussian or Voigt, the output is compared to the intrinsic wavelength of the line to estimate the velocities.
To these velocities, a specv\_evolution engine could be used to collect them for fitting with, e.g. an exponential function, for the average velocities against phase after peak, as done by \citet[][their fig. 4]{Fremling2018}, see Fig. \ref{fig:velexp}\footnote{This automatic approach works well for good signal-to-noise spectra with decent resolution, but visual inspection is frequently required to confirm the fitting for other data.}.

\subsection{Import external models}
\label{sec:addmodels}

Observers often search for intriguing transients with peculiar photometric or spectroscopic behavior that are not necessarily consistent with traditional explanations.
For example, superluminous supernovae (SLSNe) show luminosities significantly greater than for conventional SNe.
Other models for the powering mechanism are then needed to explain them rather than the classical radioactive models, e.g. a magnetar model \citep{Ostriker1971,Kasen2010,chatzo2012,Inserra2013,Nicholl_2017,liu2022}.
Instead of upgrading the built-in models in \haffet, we choose to provide users the ability to create their own models (see Sect.~\ref{sec:model}) or import models from other modeling-specific software packages, such as MOSFIT and redback. We show an example that fits the redback magnetar model\footnote{Developed based on MOSFIT package. We use the redback one because its code structure is similar to \haffet.} on the famous SN 2015bn in section~\ref{sec:magon15bn}.

One point we need to emphasize is that \haffet's modular design allows users to perform individual analyses and model fitting steps on SNe data, starting from raw photometric and spectral data, progressing through color, temperature, etc, all the way to the final bolometric LC. Alternatively, similar to some model-fitting programs like MOSFIT, all steps can be integrated into a closed-loop fitting process, offering users more flexibilities. Additionally, \haffet provides more fitting options, such as utilizing {\tt scipy} instead of MCMC (Markov Chain Monte Carlo) for faster parameter optimization. The user-friendly GUI interface within \haffet also offers convenience for users lacking in background knowledge.

\section{Fitting methods and outputs}
\label{sec:outputs}

In Sec. \ref{sec:fitting}, we introduced built-in engines within \haffet, which offer data for fitting diverse models. These engines, together with user-selected objective functions, systematically investigate optimal parameter distributions through various fitting procedures. In this section, we will explore the different fitting routines offered by \haffet and how it preserves certain outputs of this process for caching and sharing fitting results.

\subsection{Hyperparameter optimization routine}
\label{sec:optimize}

By assuming the observed deviations between the objective function and the observed data are the result of Gaussian scatter, the log-likelihood is:
\begin{equation}
\ln {\cal L} = - \frac{1}{2}\sum_{i=1}^{n} \frac{\left(O_i-M_i\right)^2}{{\beta\sigma_i}^2} - \sum\ln(\beta\sigma_i),
\end{equation}
where $O_i$, $\sigma_i$, and $M_i$ are the $i$-th of $n$ observed magnitudes, errors, and model magnitudes, respectively. The term $\beta$ is used to account for the fact that the data uncertainties are over- or underestimated. In \haffet, users can decide whether to include $\beta$ into likelihood functions with the \texttt{fit\_error} parameter (see Sect. \ref{sec:model}).
To optimize the parameters that are used in the fits to the analytic models there are two routines available in \haffet, i.e. optimization via {\tt scipy.optimize} and Monte Carlo (MC) methods via {\tt emcee} \citep{emcee}. 

{\tt scipy.optimize} offers functions designed for minimizing objective functions, potentially under constraints. It is a versatile tool that integrates various models and methods, capable of fitting the objective function based on user requirements. Its fitting speed is notably rapid, yet it frequently places significant reliance on prior information, specifically initial guesses and boundaries. In cases where prior information is insufficient, this method often fails to converge. Therefore, we also introduced MCMC method, a more resource-intensive yet more reliable approach, to \haffet.

In order to alleviate the computational cost as well as to speed up the MC burn-in process, by default \haffet first fit the model with minimizing objective functions via {\tt scipy.optimize}. The resulting parameters as well as the predefined boundaries are afterwards used to set the initial guesses for a number of Markov chains (walkers). For each chain, the set of parameters is randomly sampled, a likelihood is calculated between the observed data and the resulting models, and MC would enforce random sampling to follow the direction to minimize the likelihood. 
The parameter distributions are afterwards built after the sampler burn-in to a specific parameter region. 
To estimate the parameter error bars, by default we use 1 sigma for {\tt scipy} estimations and a correspondence of 16 and 84 percentiles for all burn-in samples as lower and upper limits.

While fitting the SNe data in Paper 1 with \haffet, we employed specific fitting parameters that were broadly suitable for the majority of the objects. Consequently, these parameters have been adopted as the default fitting options in \haffet. We emphasize that in cases where fits using these default settings are failed, users should consult the \haffet documentation page as well as the websites of {\tt scipy} and {\tt emcee}. This will provide a deeper insight into the underlying concepts and allow for the appropriate adjustment of parameters.
Moreover, a comprehensive comparison from multiple angles, e.g. posterior parameter distributions with {\tt scipy} and {\tt emcee}, are presented in section~\ref{sec:emceevsmin}.

\subsection{Caching Data and Models}
\label{sec:cache}

\haffet handles data and fits for various SNe on a case-by-case basis, and below we describe in detail how it operates.
First, a global variable called {\tt ZTFDATA}\footnote{\haffet access to ZTF data via ztfquery \citep{ztfquery}, that follows the IRSA data structure for data storage. In order to be consistent, \haffet adopted the same global variable and data structure not only for ZTF data but also other resources such as the ATLAS forced photometry.} has to be defined (often in the shell profiles; if not, it will use the current folder). This is the directory for \haffet to manage cached data, e.g. the photometric data downloaded from online resources.
All observational data are stored in dedicated folders, organized as easily readable text files.
In order to increase code efficiency, it is important to properly cache the samplings as model fitting may take a long time, especially when using the Monte Carlo method. 
We store and load the cached fitting samplings in {\tt hdf5} format with the joblib\footnote{\href{https://joblib.readthedocs.io/en/latest/index.html}{https://joblib.readthedocs.io/en/latest/index.html}} package, and all these cached files are stored following a dedicated data structure.
\haffet offers functions that facilitate the storage and retrieval of all pertinent observational data and models used in the fitting process. Additionally, it ensures cross-compatibility across various systems and platforms.

\begin{figure*}
\centering
    \includegraphics[width=0.8\textwidth]{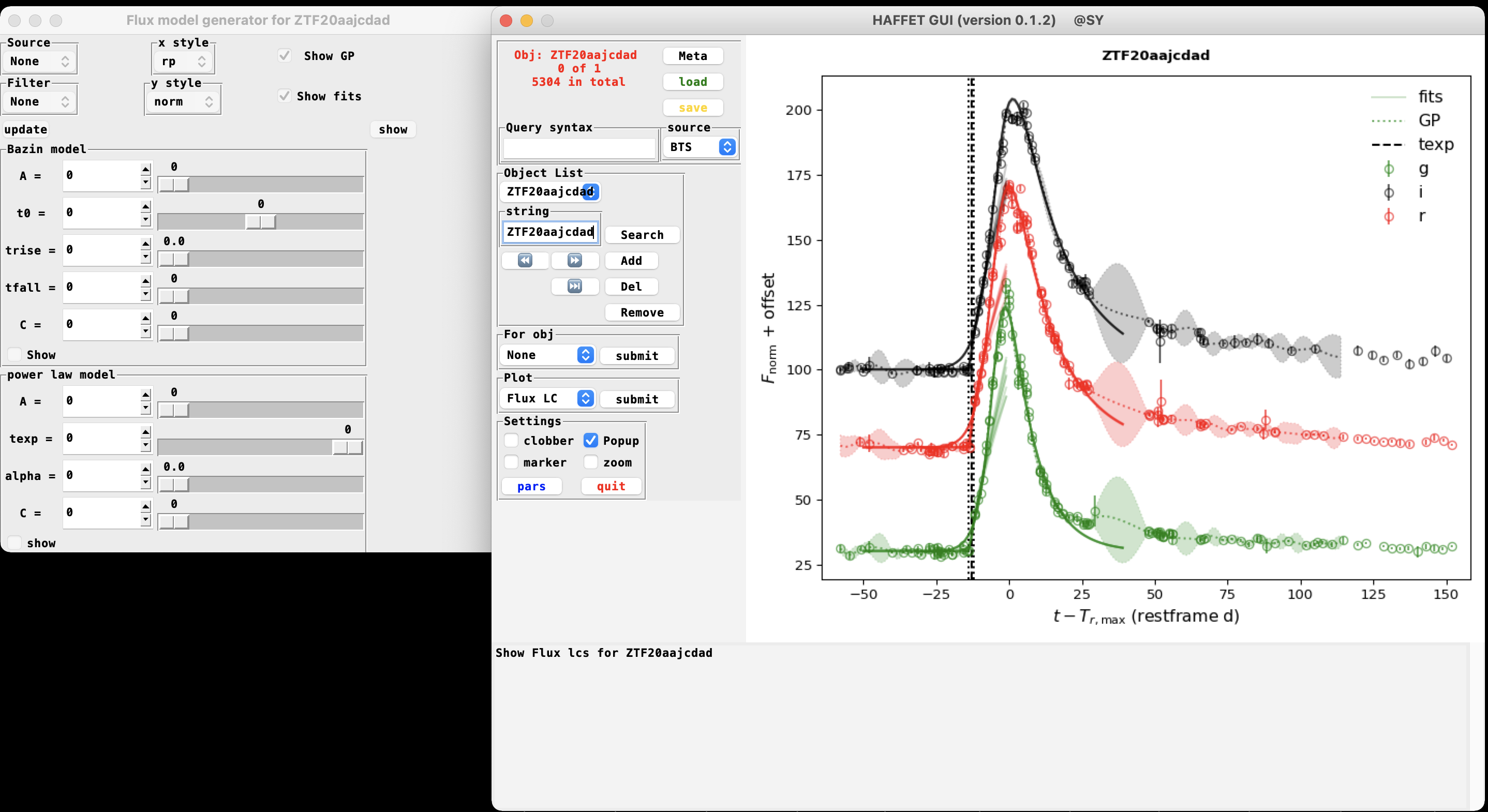}
    \includegraphics[width=0.45\textwidth]{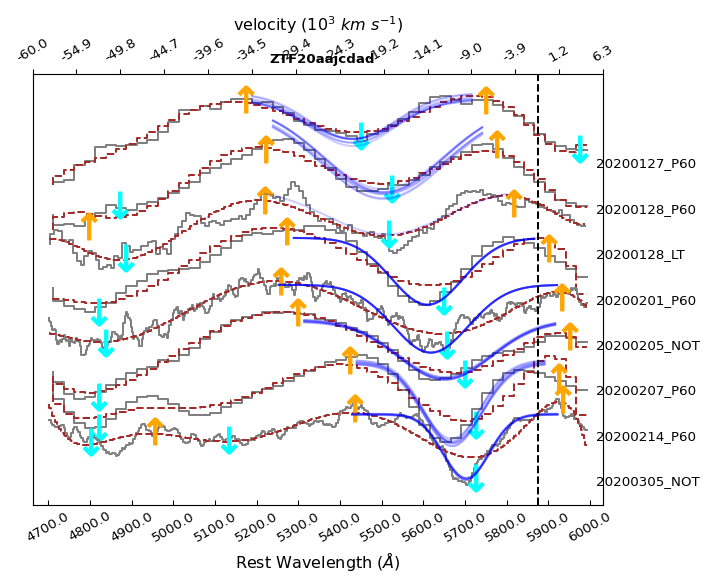}
    \includegraphics[width=0.45\textwidth]{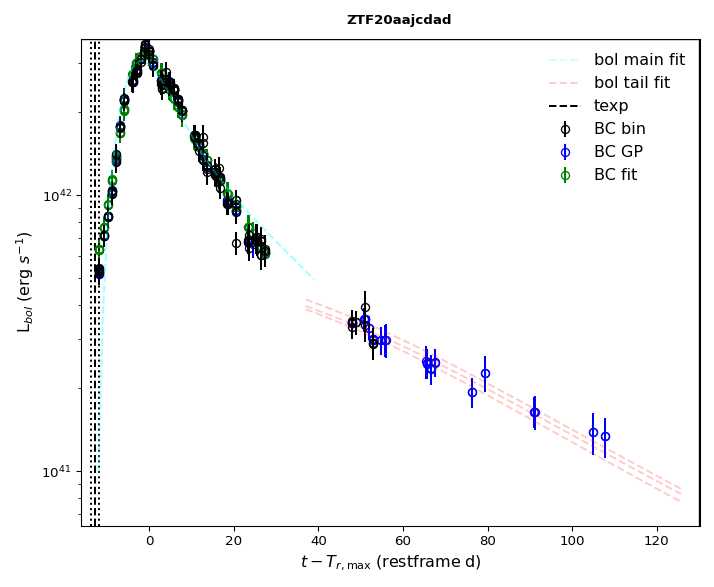}
    \caption{A view of the \haffet GUI showing the model fittings on multi-band LCs, spectral lines and the constructed bolometric LC. In the upper panel we show the control window of the GUI.}
    \label{fig:gui}
\end{figure*}

Then, according to the intrinsic properties and data qualities, different SN data may be fitted to models with different options, e.g. whether there is a tail in the SN LCs, or what is the range of photospheric phases for the Arnett model fitting, or if the automatic identification of absorption features is correct.
\haffet provides default settings in a file for general usage, as well as a possibility for the users to specify parameters for individual SNe\footnote{The parameter files can be found in the defined {\tt ZTFDATA} directory, with {\tt default\_par.txt} for the general options, and {\tt individual\_par.txt} for specific SNe.}.
An alternative option to adjust fitting options for particular SNe is to use \haffet with a Graphical User Interface (GUI), see Figure~\ref{fig:gui}. All the data, fitting parameters, and fitting samplings can be stored and loaded, which is universal to different modes.

\section{Examples of utilizing HAFFET}
\label{sec:application}

We have presented how \haffet might be used to analyze SN Ibc data in Paper I, e.g. figure 2 of Paper I illustrate the fittings with an example of SN 2020bcq; in this section, we provide some more technical descriptions as a supplement to that \haffet usage.

\subsection{Blackbody vs bolometric correction inferred luminosity}

\begin{figure*}
\centering
    \includegraphics[width=0.45\textwidth]{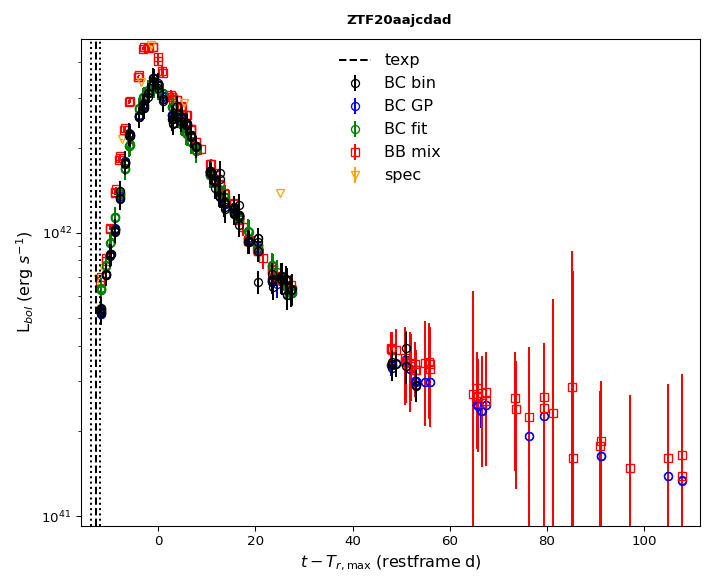}
    \includegraphics[width=0.4\textwidth]{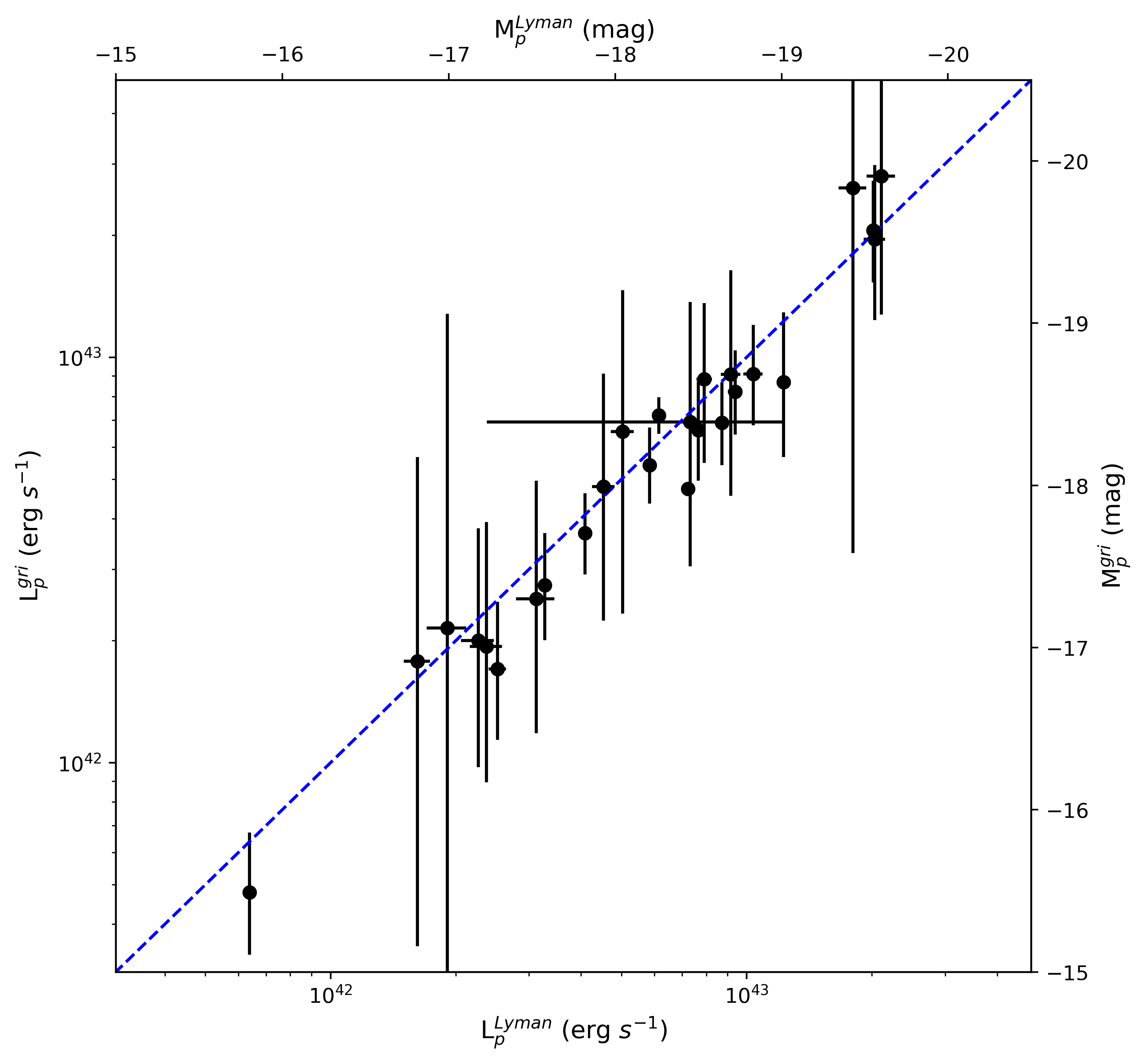}
    \caption{Luminosity comparison between different approaches. In the left panel, the circles represent the luminosity estimated with the analytic bolometric correction from \cite{Lyman14}, while the red circles are bolometric luminosities estimated from blackbody fits on the $gri$ photometry and the orange down pointing arrows are estimates based on integrating the spectra. In the right panel, we show the peak luminosity comparison between the blackbody fitted and the Lyman inferred values for 34 SNe Ib and 35 SNe Ic at peak epoch. The blue dashed line is the one-to-one relation. As shown, the luminosities from the two approaches are consistent, although the $gri$-constructed blackbody provides luminosity estimates with large(r) uncertainties.}
    \label{fig:Lpcomp}
\end{figure*}

In \haffet, we often construct bolometric LCs for SE SNe and SNe II with two bands (e.g. their $g$ and $r$ magnitudes) using the analytic bolometric corrections derived in \cite{Lyman14}.
For objects observed in more than two bands, e.g. in the ZTF $gri$ bands, 
we could also use the multi-band photometry to construct SEDs. Alternatively, we can estimate the bolometric luminosity by integrating the absolute-calibrated spectra.  We compare the luminosities estimated from these different approaches in Fig.~\ref{fig:Lpcomp}.
As shown, the bolometric correction approach using only two bands provides quite similar bolometric luminosities to those from the blackbody or spectra integration. We estimated the luminosities for ZTF SNe Ic-BL in \cite{Corsi2022} and in \cite{Anand2023}, and for comparison we test the methods also on a few famous SNe, e.g. SN 1998bw, and found consistent results between the different methods. Therefore, we set the Lyman bolometric correction method as the default approach for SE SNe and SNe II in \haffet, and provide functions to convert $BVR$ to $gr$ in case an object is observed in Johnson filters. 

\subsection{Explosion properties from the Arnett model vs the radioactive decay tail}

\begin{figure*}
\centering
    \includegraphics[width=0.3\textwidth]{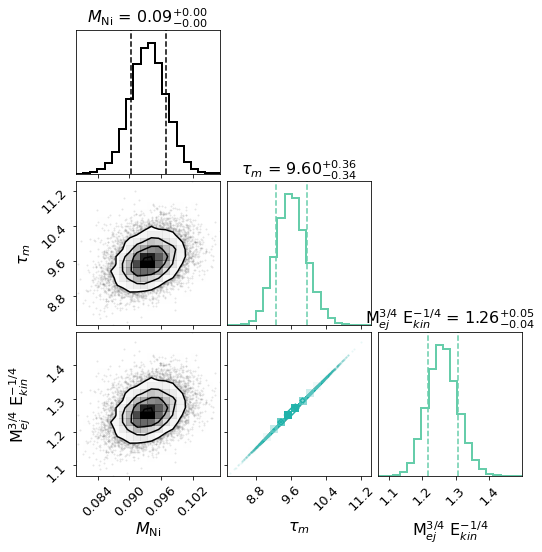}
    \includegraphics[width=0.3\textwidth]{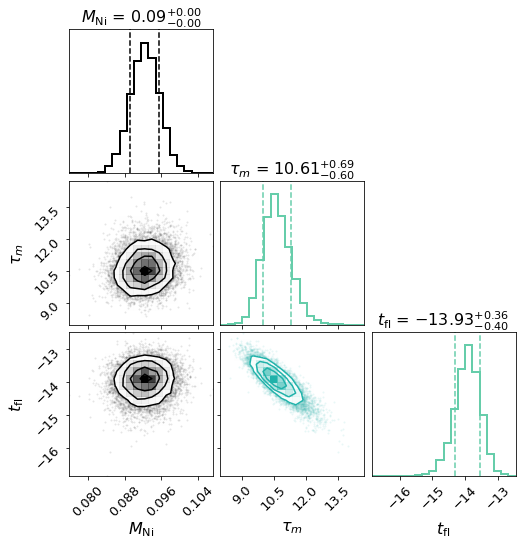}
    \includegraphics[width=0.3\textwidth]{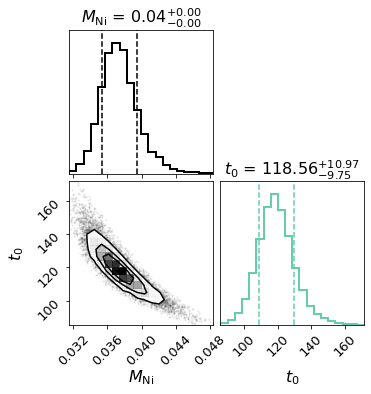}
    \caption{As shown in the lower left panel of Fig.~\ref{fig:gui}, we fit the Arnett and tail functions to the Lyman-inferred bolometric luminosity LC of SN 2020bcq (ZTF20aajcdad) as an example. The marginalized parameter distributions are shown as corner plots in this figure. In the left panel, we fit the Arnett model by using the explosion epoch from our earlier power-law fits, while in the middle plot we treat the explosion epoch as a free parameter in the Arnett model. The predicted rise time from the Arnett model, i.e. the time between the estimated explosion epoch to peak, is 13.93 days, which is comparable to the estimate from the power-law fits (13.14 days). The slight difference in the explosion epoch does not change the nickel mass estimate but modifies $\tau_m$ from 9.6 to 10.6 days. The marginalized parameter distributions of the tail fitting are shown as corner plots to the right. The best-fit nickel mass is 0.04 solar masses, which is roughly half of that inferred from the Arnett model.}
    \label{fig:arnettcompare}
\end{figure*}

\begin{figure*}
\centering
    \includegraphics[width=0.6\textwidth]{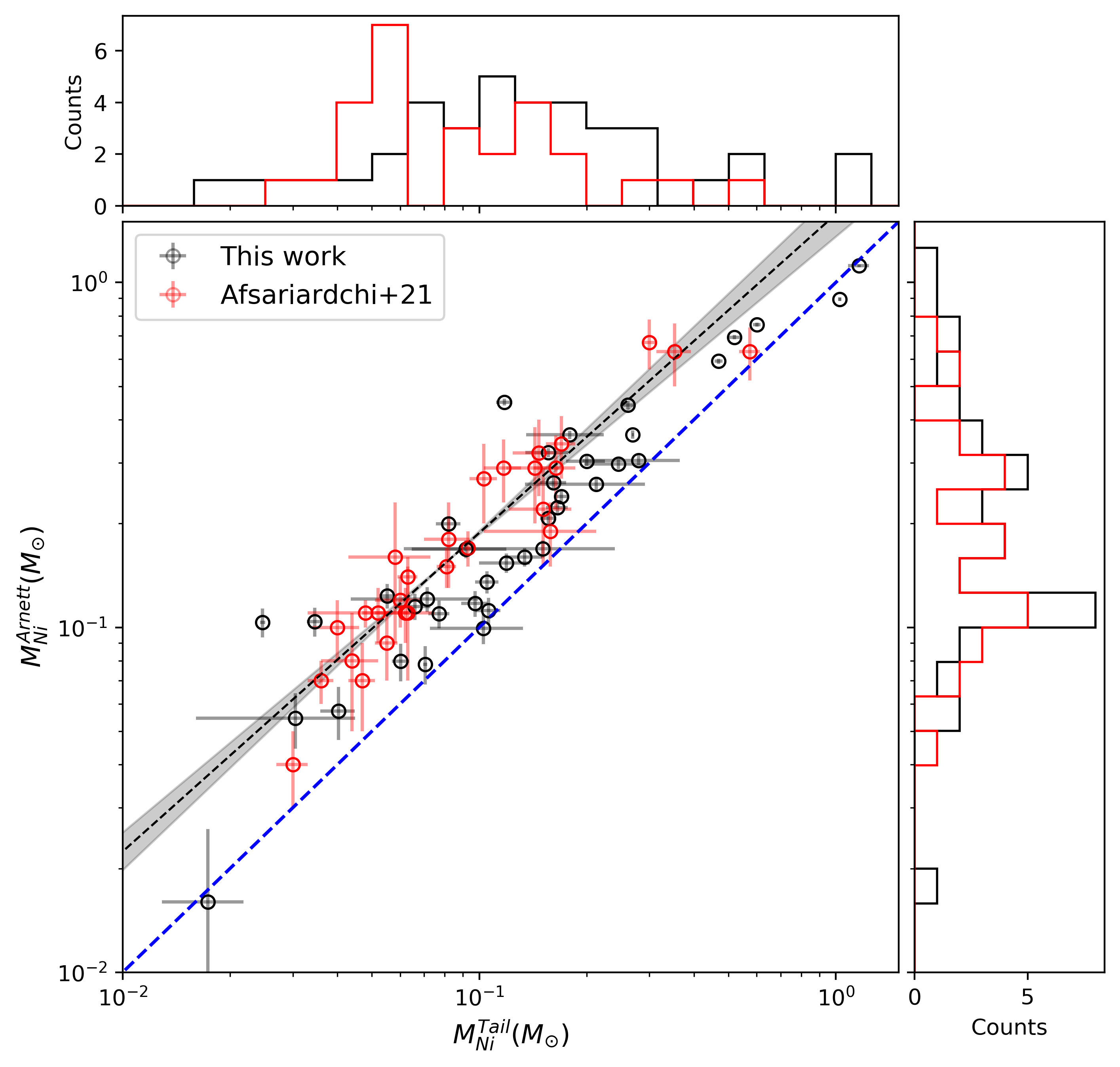}
    \caption{Comparison of \Mni~ between Arnett fits and fits to the radioactive decay tail. The ZTF SNe Ibc (the subset of SNe in Paper I that clearly shows a tail) are shown in black dots, while the sample from \cite{Afsariardchi2021} (that is a selection from the literature, including 27 SE SNe) is shown in red dots. The blue dashed line is the one-to-one relation. As shown, the Arnett inferred \Mni~ is systematically higher than those from the tails.}
    \label{fig:arnettvstail}
\end{figure*}

As discussed in section~\ref{sec:arnett}, the constructed bolometric LC at epochs around the main peak can be fitted to the classical Arnett model for an estimation on the explosion properties, e.g. \Mni, \Mej~ and \Ek. In section~\ref{sec:tail} the fits on the bolometric tail were introduced which can provide alternative estimates of the same properties. 
In Fig.~\ref{fig:arnettcompare} we compare the nickel masses from the two approaches with SN 2020bcq as example. As shown, the inferred \Mni~ from the Arnett model is 0.09 solar masses, which is $\sim 2$ times larger than the estimate from the tail that is 0.04 solar masses. 
We conducted such a comparison to a sample of ZTF SNe Ibc (the black dots) in Fig.~\ref{fig:arnettvstail} and conclude similarly to \cite{Afsariardchi2021} (the red dots, see also their Fig. 4) that the Arnett method yields \Mni~ values which are systematically a factor of around 2 larger than those estimated from the tail. 

The restrictions of the Arnett model may be to blame for this discrepancy, e.g. \cite{KK19} claimed that the self-similarity condition breaks down when the SN ejecta have a centralized \Ni~ deposition, thus proposed a new analytical model for SE SNe that introduced a dimensionless parameter $\beta$ to the Arnett model. 
A robust $\beta$ for each SE SN sub-type is still under investigation, e.g. \cite{Meza2020} adopted $\beta$ values that were derived from the simulated SE SN LCs of \cite{Dessart2016}, and \cite{Afsariardchi2021} instead derived $\beta$ from a sample of literature SE SNe.  As the largest untargeted Type Ibc survey to date, the ZTF BTS offers a large amount of SE SNe for calibrating $\beta$. 

\begin{figure*}
\centering
    \includegraphics[width=0.6\textwidth]{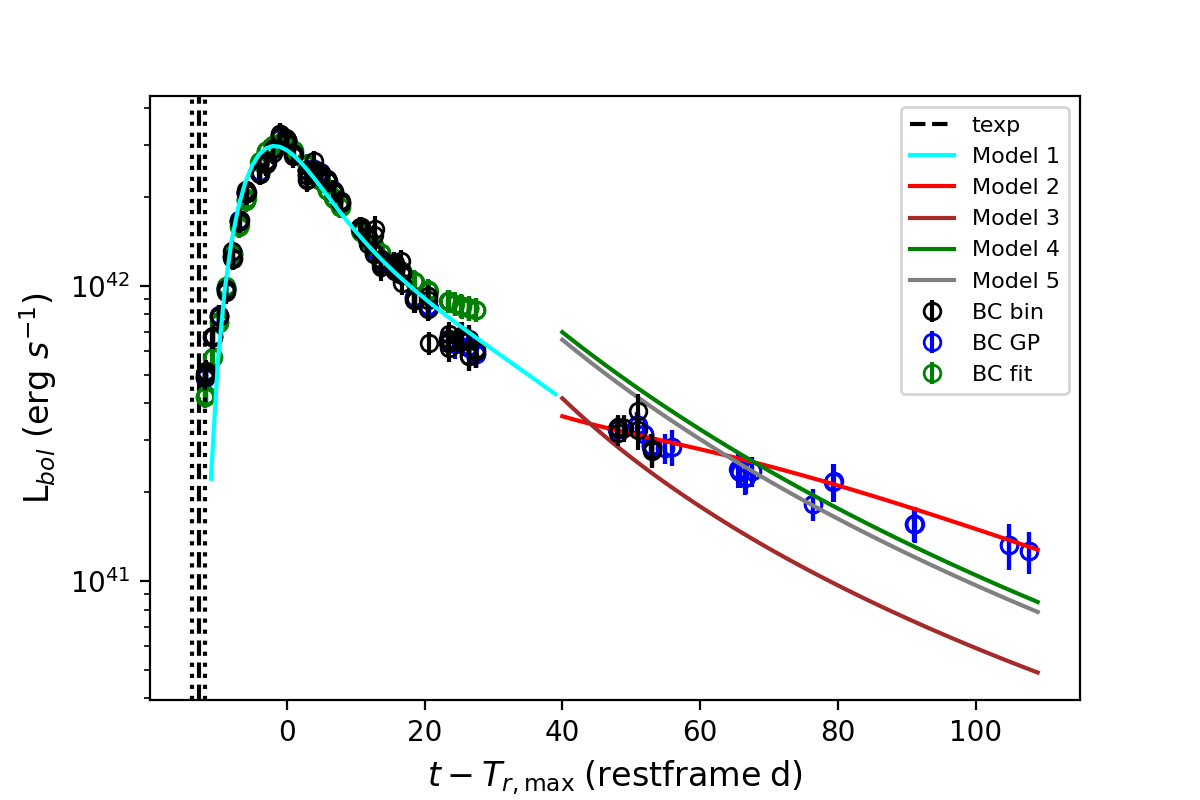}
    \caption{Various models fit to the bolometric LC of SN 2020bcq (the open circles). The realization with the best-fit Arnett and tail models are shown by models 1 and 2 (the solid cyan and red lines; see also the lower right panel of Fig. \ref{fig:gui}). From model 1, we obtain the posterior distributions of \Mni~ and $\tau_m$, which can be translated into \Mej~ and \Ek~ together with a velocity measurement ($\sim$ 10,000 $\rm km~s^{-1}$). We use the parameters obtained from model 1 to reproduce a tail bolometric, model 3 (the brown line). We adapted its $\gamma$ trapping opacity (model 4, $\rm \kappa_\gamma$ from 0.027 to 0.05 $\rm cm^2g^{-1}$) or photospheric velocity (model 5, $V_{\rm{ph}}$ from 10,000 to 5,000 $\rm km~s^{-1}$) so that the tail can be brighter to match the observed data.}
    \label{fig:arnetttail}
\end{figure*}

In Fig.~\ref{fig:arnetttail}, we explore whether explosion parameters from the radioactive tail could be consistent with the peak Arnett fit. 
Models 1 and 2 are the best-fit models for the Arnett peak and bolometric tail, respectively,  as estimated using \haffet with constant opacities, i.e. $\kappa=0.2~{\rm cm^2g^{-1}}$ and $\rm \kappa_\gamma=0.027~cm^2g^{-1}$. Assuming the Arnett model is accurate, the bolometric tail could be predicted (model 3) using its inferred explosion parameters. As shown, the radioactive decay model, i.e. model 3, is underestimated compared to the Arnett model. It is possible to argue that the adopted velocity (that was estimated via the Dessart relation, see equations \ref{eq:mej} and \ref{eq:ek}) or the opacities are incorrect. We then explore how the tail brightness and slope depend on the velocity and opacities in models 4 and 5. 
We can translate the Arnett model's estimates of $\tau_m$ into \Mej~ and \Ek~ (along with a velocity measurement), and then $T_0$ for the tail.
In model 4, we fix the \Mni, $\tau_m$, and fit the bolometric tail with $\gamma$-trapping opacity as a free parameter, whereas in model 5 we fit the velocity instead\footnote{For hyperparameter optimization, we use the MCMC routine in \haffet with the likelihood containing term $\beta$, see section \ref{sec:optimize}.}.
As shown, the best fits of models 4 and 5 have steeper slopes than model 2.

\subsection{Optimize fitting with minimize routine vs monte carlo}
\label{sec:emceevsmin}

\begin{figure*}
\centering
    \includegraphics[width=0.6\textwidth]{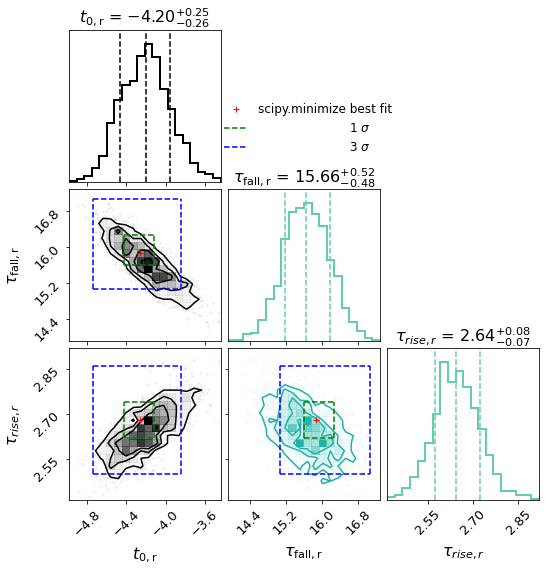}
    \caption{As shown in the upper panel of Fig. \ref{fig:gui}, we fit the multi-band LCs following analytic models of \cite{bazin} with SN 2020bcq (ZTF20aajcdad) as an example. 
    Such fits can be done in different routines, and in this plot we compare the fitting results between them. The parameter distributions from MCMC on $r$-band LC are shown as corner plots. 
    The vertical dashed lines in the marginalized parameter histograms represent the 16 and 84 percentile of all burn-in samples, which are corresponding to the 1 sigma posterior parameter distributions from the {\tt scipy} minimize routine (the green box).}
    \label{fig:fitcompare}
\end{figure*}

As discussed in section \ref{sec:optimize}, there are two routines available in \haffet for hyperparameter optimization, and in this section we compare them for the fitting results on the $r$-band LC of SN 2020bcq with the analytic model of \cite{bazin}. 
The parameter contours from the MCMC and the fitted region from the {\tt scipy} minimization are shown in Fig.~\ref{fig:fitcompare}. 
Once fit with the {\tt scipy} package, the optimal values for the parameters as well as their estimated covariances could be estimated. In \haffet, the diagonals of the covariance matrix are used as the errors for the parameters, therefore hyperparameters could be optimized into the green and blue dashed regions (for 1 and 3 sigma correspondingly) as shown in Fig.~\ref{fig:fitcompare}. 
In \haffet, we mimic the MCMC contours for the {\tt scipy} minimization routine by randomly selecting parameters within the permitted regions.
Also, the resulting parameter ranges from {\tt scipy} minimize, i.e. the blue and green dashed regions, can afterwards be used as prior for MC samplers. After burn-in, a more degenerate version of the contours could be generated from the MCMC process. 
To decide the parameter error bars, in \haffet we use 1 sigma for the {\tt scipy} estimations, and corresponding 16 and 84 percentiles of all burn-in samples as lower and upper limits.

As shown in Fig.~\ref{fig:fitcompare}, the {\tt scipy} minimize routine can estimate a similar set of parameters as MCMC, with much less computational cost and time. However, the {\tt scipy} method might not be suitable for complicated models with many free parameters, and for such cases MCMC is the better option.

\subsection{Beyond the Arnett model, a magnetar fit to SN 2015bn}
\label{sec:magon15bn}

\begin{figure*}
\centering
    \includegraphics[width=0.45\textwidth]{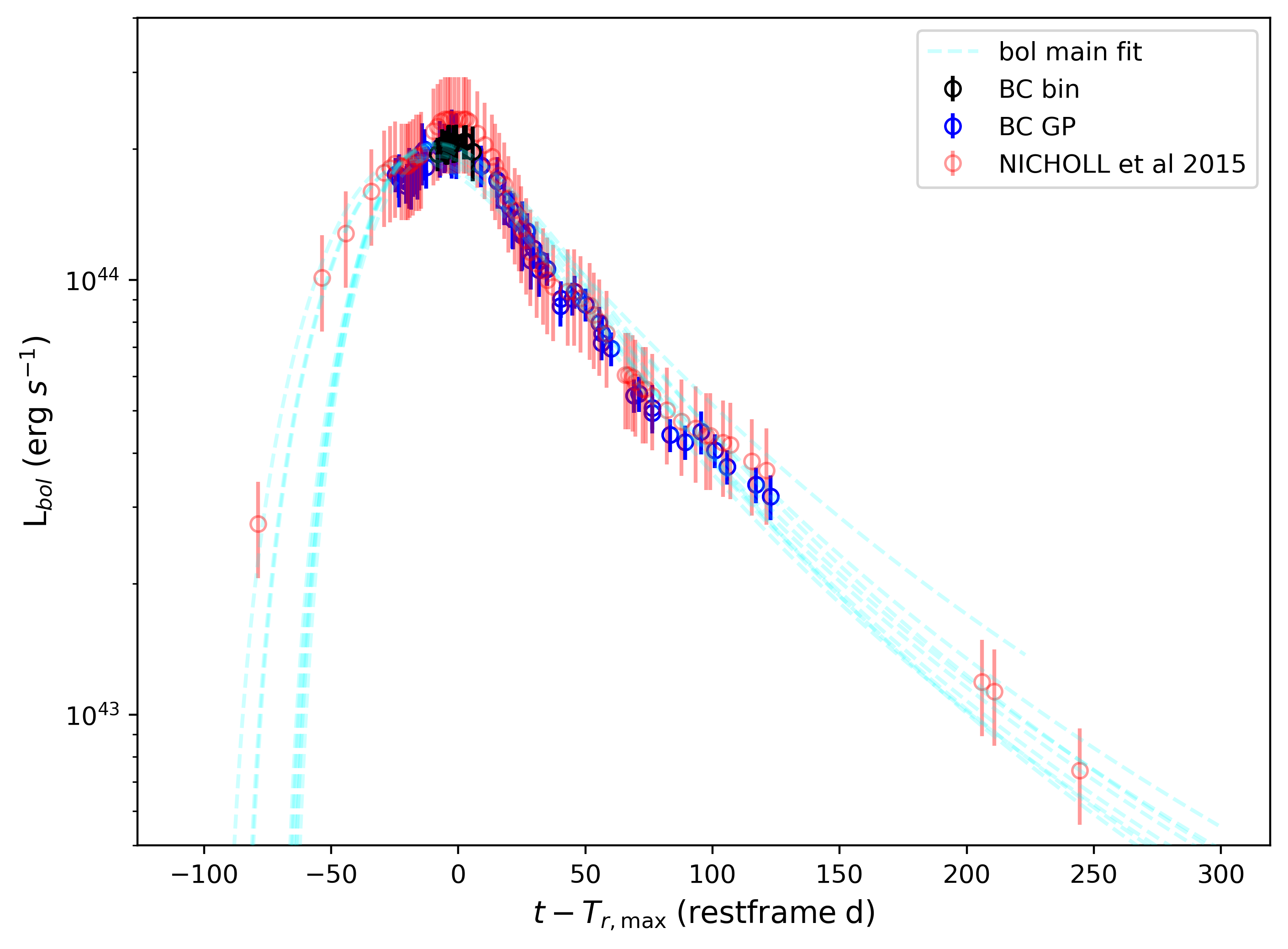}
    \includegraphics[width=0.4\textwidth]{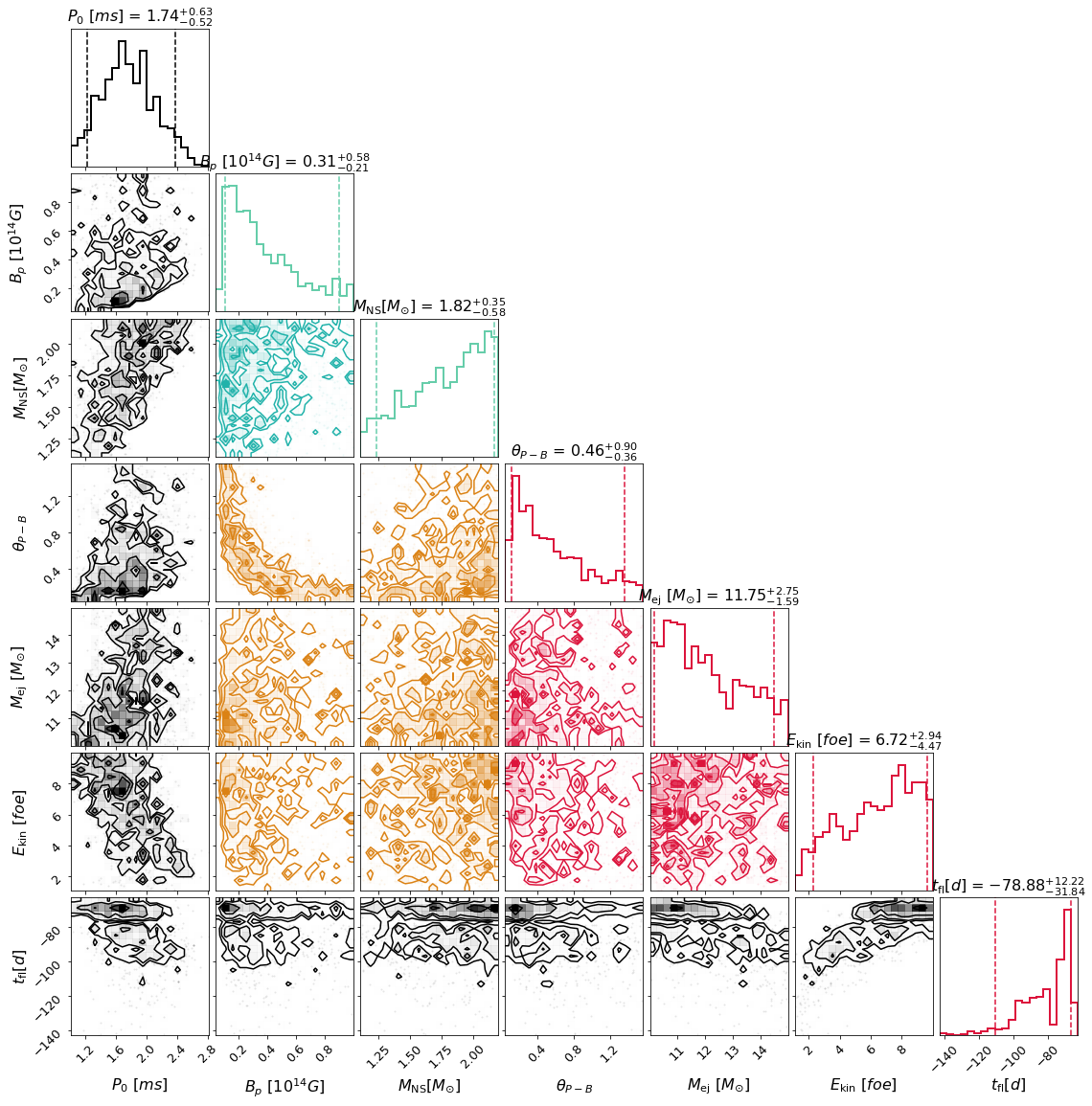}
    \caption{Magnetar model fitting on SN 2015bn via \haffet. We construct the bolometric LC of SN 2015bn with $g$ and $r$ magnitudes with the analytic bolometric correction for SE SNe derived in \cite{Lyman14}. They are shown in black and blue open circles. As shown, they are quite similar to the one from \cite{2015bn}, i.e. the red circles. The cyan dashed lines are reproduced with the fitted magnetar model samplings. The parameter contours are shown to the right, which is similar to the estimates of \citet[][their Fig. 4]{2015bn}.}
    \label{fig:2015bn}
\end{figure*}

As discussed in section \ref{sec:addmodels}, \haffet can be used to fit the constructed bolometric LC to models from external packages. In this section, we show an example with a magnetar model (import from MOSFIT) fitting on data of SN 2015bn \citep{2015bn}. The photometric data were obtained from the Open SN Catalog \citep{oac}.
We use the analytic bolometric correction for SE SNe derived in \cite{Lyman14} to estimate its bolometric LC. 
As shown in the left panel of Fig. \ref{fig:2015bn}, the bolometric LCs for SN 2015bn from \cite{2015bn} and \haffet are similar, however the \haffet LC lack a few early phase observations since they were not available in the Open SN Catalog. The absence of early phase constraints gave more freedom to our fittings, see the early stage of the cyan dashed lines that are the modeled magnetar LCs fitting on the \haffet bolometric LC. For the magnetar fitting, we set the same host extinction and opacities for SN 2015bn as in \cite{2015bn} and fit the explosion epoch as a free parameter. The resulting contours are presented in the right panel, which is consistent with those from \cite{2015bn}. 

\section{Summary and Future prospects}
\label{sec:conclusion}

In this work we have introduced a new open-source SN fitting code \haffet. We explain the driving forces behind the \haffet code's creation, and how \haffet could be utilized to analyze SN data for sample investigation. 
We tested \haffet by applying it to the SNe Ibc that were observed by ZTF and classified with BTS in Paper I. In this paper we show more examples and demonstrate that \haffet is functional and clearly not strictly adhering to SNe Ibc. 

To broaden its applicability, we have plans for future upgrades of \haffet. This includes:

\begin{enumerate}

  \item Including routines to query data from more open online resources, e.g. alert brokers such as Lasair\footnote{\url{https://lasair-ztf.lsst.ac.uk/}} and Alerce\footnote{\url{https://alerce.science/}}.
  
  \item K-corrections for high-redshift SNe, which would be important in the LSST era.
  
  \item Spectroscopic classifications via, e.g. \cite{NGSF}. 
   
  \item Besides analytic models, we intend to provide possibilities for users to include hydrodynamic models as well. This could be done by comparing data to a grid of hydrodynamic model realizations.

  \item Comprehensively improving the radioactive models, e.g. by including nickel mixing \citep{Dessart:2012aa, Dessart2016, Yoon2019}, asymmetry, etc. These can be explored after a grid of numerical models are available, which are not possible in the analytic Arnett model.

  \item Adding machine learning components to {\tt snelist} for sample exploration. For instance a dataset of SN LCs, \haffet can be used to obtain a list of features such as their peak magnitudes, slopes, and colors. A Principal Component Analysis could be then adopted to asses which properties are significantly related to the SN types.  
\end{enumerate}

\section*{Data availability}

The codes and data underlying this article are available in
\url{https://github.com/saberyoung/HAFFET}

\software{numpy \citep{harris20a},
          matplotlib \citep{hunter07a},
          scipy \citep{virtanen20a},
          pandas \citep{mckinney10a},
          emcee \citep{emcee},
          astropy \citep{astropy:2013, astropy:2018, astropy:2022},
          george \citep{george}.
          }

\begin{acknowledgements}
We thank Gokul Prem Srinivasaragavan, Kaustav Kashyap Das, Ping Chen, Xiaofei Dong, Songyu Shen for their help testing \haffet codes in their devices, and Matt Nicholl, Enrico Cappellaro, Conor Michael Bruce Omand, Nikhil Sarin, Avishay Gal-Yam, Mansi Kasliwal, He Gao and Liangduan Liu for helpful comments.
Based on observations obtained with the Samuel Oschin Telescope 48-inch and the 60-inch Telescope at the Palomar Observatory as part of the Zwicky Transient Facility project. 
ZTF is supported by the National Science Foundation under Grant No. AST-1440341 and a collaboration including Caltech, IPAC, the Weizmann Institute for Science, the Oskar Klein Center at Stockholm University, the University of Maryland, the University of Washington, Deutsches Elektronen- Synchrotron and Humboldt University, Los Alamos National Laboratories, the TANGO Consortium of Taiwan, the University of Wisconsin at Milwaukee, and Lawrence Berkeley National Laboratories. Operations are conducted by COO, IPAC, and UW. 
This work was supported by the GROWTH project \citep{Kasliwal2019}
funded by the National Science Foundation under PIRE Grant No 1545949. 
The Oskar Klein Centre was funded by the Swedish Research Council.
Gravitational Radiation and Electromagnetic Astrophysical Transients (GREAT) is funded by the Swedish
Research council (VR) under Dnr 2016-06012. 
Partially based on observations made with the Nordic Optical Telescope, operated by the Nordic Optical Telescope Scientific Association at the Observatorio del Roque de los Muchachos, La Palma, Spain, of the Instituto de Astrofisica de Canarias. Some of the data presented here were obtained with ALFOSC. 
The SED Machine is based upon work supported by the National Science Foundation under Grant No. 1106171. 
The ZTF forced-photometry service was funded under the Heising-Simons Foundation grant \#12540303 (PI: Graham).
This work has been supported by the research project grant “Understanding the Dynamic Universe” funded by the Knut and Alice Wallenberg Foundation under Dnr KAW 2018.0067. 

\end{acknowledgements}

\bibliography{ref}

\end{document}